\documentclass{article}

\usepackage{amsmath,amssymb}
\usepackage{lmodern}
\usepackage{iftex}
\usepackage{graphicx}  % standard LaTeX graphics tool;
\usepackage{amsmath}   % For getting proper math eqns
\usepackage{amssymb}   % Used for getting various symbols eg. gtrsim, lesssim etc.
\usepackage{bm} % For bold math (esp of lower greek letters)
\usepackage{dcolumn}% Align table columns on decimal point
\usepackage{color}
\usepackage{mathrsfs}
\usepackage{amsfonts}
\usepackage{varioref}
\usepackage[margin=1.6in]{geometry}
\usepackage{slashed}
\usepackage{authblk}
\usepackage{float}
\usepackage{subcaption}
\restylefloat{figure}
\RequirePackage[colorlinks,citecolor=blue,urlcolor=magenta,linkcolor=blue]{hyperref}
\def\p{\partial}
\def\e{\epsilon}

\def\be{\begin{equation}}
\def\ee{\end{equation}}

\labelformat{section}{Section #1} 
\labelformat{subsection}{Section #1} 
\labelformat{subsubsection}{Section #1}
\labelformat{subsubsubsection}{Section #1}
\labelformat{equation}{Eq.~(#1)} 
\labelformat{figure}{Fig.~#1} 
\labelformat{subfigure}{Fig.~\thefigure#1} 
\labelformat{table}{Tab.~#1} 
\labelformat{appendix}{Appendix #1}
\title{\bf Self interacting scalar field theory  in  general curved spacetimes at zero and finite temperature revisited}
\author[]{Vishal Nath\footnote{vishaln.physics.rs@jadavpuruniversity.in} \, and \, Sourav Bhattacharya\footnote{sbhatta.physics@jadavpuruniversity.in}}
\affil[]{Relativity and Cosmology Research Centre, Department of Physics, Jadavpur University,\\ Kolkata 700 032, India}

\begin{document}
\maketitle
\begin{abstract}
\noindent
We revisit  the problem of spontaneous symmetry breaking (SSB), its restoration, and phase transition for a self interacting quantum scalar field  in a general curved background, at zero and finite temperature. To the best of our knowledge, most of the earlier computations  in this context have been done in the linear order in curvature, which may not be very suitable for the Ricci flat spacetimes. One of our objectives is to see whether  the higher order terms can bring in qualitatively new physical effects, and thereby attempting to fill in this gap in the literature. We use Bunch and Parker's local momentum space representation of the Schwinger-DeWitt expansion of the Feynman propagator. Such expansion, being based upon the local Lorentz symmetry of spacetime, essentially probes the leading curvature correction to short scale, ultraviolet quantum processes. We compute the renormalised, background spacetime curvature (up to quadratic order) and temperature dependent one loop effective potential for $\phi^4$ plus $\phi^3$ self interaction.    In particular for the de Sitter spacetime, we have shown for the $\phi^4$-theory that we can have SSB  even with a positive rest mass squared and positive non-minimal coupling, at zero temperature.   This cannot be achieved by the linear curvature term alone and the result remains valid for a very large range of renormalisation scale. Such  SSB will generate a field mass that depends upon the spacetime curvature as well as the non-minimal coupling.    For a phase transition, we have  computed the leading curvature correction to the critical temperature.  At finite temperature, symmetry restoration is demonstrated.  We also extend some of the above results to two loop level. The symmetry breaking in de Sitter at two loop  remains  present. We have further motivated the necessity of treating this problem non-perturbatively in some instances. 
\end{abstract}
\vskip .5cm

\noindent
{\bf Keywords :} Thermal field theory, curved spacetime, effective potential, symmetry breaking, de Sitter

\pagebreak
\tableofcontents
\section{Introduction}\label{S1}
%%%%%%%%%%

\noindent
The motivation of studying  quantum field theory in curved backgrounds is to investigate the effects of spacetime curvature or geometry on the fluctuations or propagation of quantum fields living in it. This much cultivated yet still active subject, originally  supposed to be a prelude to a full theory of quantum gravity,  has led to interesting physical predictions like the particle creation in cosmological or black hole backgrounds, fluctuations in the  early inflationary universe, regularisation of the stress energy momentum tensor and so on, over decades. We refer our reader to~\cite{Dewitt, Birrell:1982ix, Khlopov, Anto, Toms} and references therein for  a vast historic overview.

We are interested in looking into a self-interacting scalar field theory at zero as well as  at finite temperature
 in a general, regular curved background  in this paper.  Thermal fields in a curved background may exist in various different scenarios. For instance, the thermal fields living in the early  universe, or the hot plasma residing in a star or accreting onto a black hole, the blackbody radiation emitted by a black hole, and so on.

An important aspect of quantum theory of a scalar field is the existence of an effective action and a corresponding effective potential. Such potential arises due to the integration over the quantum fluctuations and it adds up to the tree level self interaction potential.  The total effective potential  (i.e., the tree plus the loop correction) can be qualitatively different in shape from that of the tree level. Such effective potential may lead to the highly interesting physical phenomenon of spontaneous symmetry breaking, developed in the seminal works in~\cite{Coleman, Jackiw, dolan, Wein}.  We further refer our reader to~\cite{Lee:1974fj, Ken, Fraser, Aitch, Ford, yang, Martin1, herr} and references therein for various aspects associated with symmetry breaking and effective potentials. Computation of  effective actions and potentials for field theories at finite temperature is also a physically well motivated problem e.g.~\cite{dolan,  Barry, Rod, Paul, Tanguy,  Rajesh, Moss, Pole, Guang, Zel,   Arjun, heymans, mooij}. Interestingly, the effective potential at high temperature ceases to have any symmetry breaking feature even if at zero temperature it has so, known as the symmetry restoration phenomenon.  We further refer our reader to~\cite{kapusta, Ashok, Nastase:1970yyp} and references therein for vast discussion on these issues.

We do not have the scope here to review in detail the developments of quantum field theory in curved spacetimes, and the effective action formalism. We refer our reader to~\cite{Drummond:1979pp, Bunch:1980bs, Shore, bunch, Panangaden:1981qj, Ishi, Parker, Odintsov, Jack:1983sk, Connor, Buch,   Odintsov:1988wz, Odintsov:1988ec, Buchbinder:1989bt, Odin, Inagaki:1993ya, kir, Elizalde:1993ew,  Elizalde:1993qh,  Elizalde:1994gv,  Inagaki:2005qp, Shap, Shapiro, Arai:2012sh,   LopezNacir:2013alw, Jtoms, Bhattacharya:2015hva, Mac, dePaulaNetto:2016voj, Arttu, dtoms, Toms:2019erd, Susobhan, panda} and references therein for some aspects of propagators, renormalisation, effective action and symmetry breaking issues at zero temperature. For thermal propagators and  finite temperature effects in flat and curved spacetimes, we refer our reader to~\cite{Christensen:1977jc, Candelas:1980zt, Page:1982fm, blhu, Hu:1982ue, naka, Hu, dow, Cog, DeNardo:1997gn, Balbinot:1999vg, Hat, Hay, Buchholz:2006iv, Kalinichenko, kwan, Chang:2019ebx, Khakimov:2023emy} and also references therein. As far as symmetry breaking and phase transition in a general curved background is concerned, most of the earlier works use linear in the curvature approximation and then employ some renormalisation group technique for resummation (e.g.~\cite{Ishi, Odintsov, Buch, Odin} and also references therein). For example, in~\cite{Buch},  a renormalisation group derivation of the effective potential for a massless scalar coupled to gauge field, and a discussion on phase transition can be seen. Even though this is a powerful method, it does not work for the Ricci flat spacetimes. In particular, the $ R \phi^2$-type non-minimal coupling does not contribute to the effective potential for such spacetimes.   At finite temperature on the other hand, the Page approximation technique can be useful for static spacetimes~\cite{Page:1982fm}.  To the best of our knowledge, the thermal effective Lagrangian density in a general curved spacetime for  $\phi^4$ theory, up to quadratic order in the curvature was first derived in~\cite{Hu}, by employing a proper time expansion technique.   \\

\noindent
We wish to revisit in this paper the issue of the effective potential for a self interacting scalar field theory at both zero as well as finite temperature. We wish to put an emphasis on the spontaneous symmetry breaking and its restoration and phase transition phenomena. We shall keep curvature terms up to quadratic order containing the Kretschmann scalar  and hence our result will also be valid for the Ricci flat spacetimes. One of our objectives behind this work is to see whether the higher order curvature terms could bring in qualitatively new effects, and thereby attempting to fill  in the gap that seems to be present in the literature. We shall also discuss generalisation of some of the one loop results to two loop level. 
 
The rest of the paper is organised as follows. In the next section, we  discuss very briefly the basic technical framework we work in. In \ref{S3}, we briefly derive the one loop effective potential at zero temperature, up to the quadratic order of the spacetime curvature for quartic and cubic self interactions. In \ref{S4}, we extend these results to finite temperature, with high temperature approximation. Although as we mentioned, the above problem was first addressed in~\cite{Hu}, our notation and the choice of the renormalisation scale will be a bit different. Due to this reason, we shall keep briefly the computations, for the sake of completeness and clarity. We discuss spontaneous symmetry breaking, its restoration and phase transition at zero and finite temperatures in \ref{S6}, \ref{S8}. We next generalise some of these results to two loop in \ref{S9}. In particular, the contribution from the two loop double bubble diagram under local approximation to the effective potential, at finite temperature  has been explicitly presented, \ref{S11}. Finally, we conclude in \ref{concl} with some future motivation for non-perturbative calculations. For the de Sitter spacetime in particular, we have shown that we can have symmetry breaking for a scalar even with positive rest mass squared {\it and} a positive non-minimal coupling, at zero temperature, at both one and two loop level. This cannot be achieved by the  linear curvature term alone and the result remains valid for a very large range of renormalisation scale. \\

We will work with the mostly positive signature of the metric in $d=4-\e$ ($\e =0^+$) and will set $c=\hbar=k_{B}=1$ throughout.

%%%%%%%%%%
\section{The basic set up}\label{S2}\
%%%%%%%%%%%%%

\noindent
We wish to begin by  briefly reviewing below the basic framework we will be working in, referring our reader to~\cite{Dewitt, Toms, Panangaden:1981qj, Parker} and references therein for detail. The action of the self interacting scalar field theory we are interested in reads in a curved spacetime background
\begin{eqnarray}
S = -\int d^d x\sqrt{-g} \left[\frac12 g^{\mu\nu}(\nabla_{\mu} \Phi)(\nabla_{\nu} \Phi) +\frac12 m^2 \Phi^2 +\frac{\eta\Phi^3}{3!}+\frac{\lambda\Phi^4}{4!} + \frac12\xi R \Phi^2\right] \qquad \qquad (\lambda >0).
\label{v1}
\end{eqnarray}
The corresponding equation of motion for $\Phi$ reads
\begin{eqnarray}
\left(-\square +m^2+\frac{\eta\Phi}{2} +\frac{\lambda\Phi^2}{3!} +\xi R\right)\Phi = 0.
\label{v2}
\end{eqnarray}
We are interested to do QFT in a neighborhood of a given spacetime point. Accordingly, we expand the metric in the Riemann normal coordinate ($y^{\mu}$) around it (see e.g.,~\cite{Aga} for technical detail and also for some original references)
\begin{eqnarray}
&& g_{\mu\nu} = \eta_{\mu\nu} -\frac13 R_{\mu\alpha\nu\beta} y^{\alpha} y^{\beta} - \frac16 R_{\mu\alpha\nu\beta;\gamma} y^{\alpha} y^{\beta}y^{\gamma} +\left(-\frac{1}{20}R_{\mu\alpha\nu\beta;\gamma\delta} + \frac{2}{45}R_{\alpha\mu\beta\lambda}R^{\lambda}{}_{\gamma\nu\delta}\right)y^{\alpha} y^{\beta}y^{\gamma}y^{\delta}\cdots , \nonumber\\
&& |g| = 1- \frac13 R_{\alpha\beta}y^\alpha y^\beta -\frac16 R_{\alpha\beta;\gamma}y^{\alpha} y^{\beta}y^{\gamma} +\left(\frac{1}{18} R_{\alpha\beta} R_{\gamma\delta} -\frac{1}{90} R^{\lambda}{}_{\alpha\beta}\,^{\kappa} R_{\lambda\gamma\delta\kappa} - \frac{1}{20} R_{\alpha\beta ; \gamma\delta}  \right)y^{\alpha} y^{\beta}y^{\gamma}y^{\delta}\cdots .
\label{v4}
\end{eqnarray}
The curvature terms appearing above serve as expansion coefficients, evaluated at the origin of the normal coordinate, where $y=0$. The above expansion only holds good if we are interested in a length scale small compared to the characteristic scale associated with the spacetime $\sim R^{-1/2}$, over which the variation of the curvature is effective.

Let $G(x,x')$ be the Feynman propagator in this spacetime. For computational convenience, one defines  $\bar{G}(x,x^{\prime})$ as
\begin{eqnarray}
G(x,x^{\prime}) = g^{-{\frac{1}{4}}}(x)\bar{G}(x,x^{\prime})g^{-{\frac{1}{4}}}(x^{\prime}).
\label{v5}
\end{eqnarray}
One can next use   the local flatness of the spacetime, \ref{v4}, in order to define a local momentum space, leading to a Fourier decomposition~\cite{Toms, Parker}
\begin{eqnarray}
\bar{G}(x,x^{\prime}) = \int \frac{d^d x}{(2\pi)^d}e^{ik\cdot y} \bar{G}(k).
\label{v6}
\end{eqnarray}
We now expand  $\bar{G}(k)$ as
\begin{eqnarray}
\bar{G}(k) = \bar{G_0}(k) +\bar{G_1}(k) +\bar{G_2}(k) +\bar{G_3}(k) +\cdots .
\label{v7}
\end{eqnarray}
where $\bar{G_i} (k)$ contains the $i^{\rm th}$ derivative of the metric at the centre of the Riemann normal coordinate, $y = 0$.\\

We have at the leading order
\begin{eqnarray}
\bar{G_0}(k) =\frac{1}{k^2 +m^2}
\label{v8}
\end{eqnarray}
Since there is no first order derivative of the metric surviving at $y = 0$, we have $\bar{G_1}(k) =0$. One also computes
\begin{eqnarray}
\bar{G_2}(k) =\frac{\left(\frac16 -\xi\right)R}{(k^2 +m^2)^2}, \qquad \qquad 
\bar{G_3}(k) =\frac{i\left(\frac16 -\xi\right)(\nabla_{\alpha} R)}{k^2 +m^2} \p^{\alpha}_{k}\frac{1}{k^2 +m^2}, \nonumber\\
\bar{G_4}(k) =\frac{\left(\frac16 -\xi\right)^2 {R^2}}{(k^2 +m^2)^3} +\frac{a_{\alpha\beta}}{k^2 +m^2}\p^{\alpha}_k\p^{\beta}_k\frac{1}{k^2 +m^2},
\label{v9}
\end{eqnarray}
where we have abbreviated 
\begin{eqnarray}
a_{\alpha\beta} =\frac12\left(\xi -\frac16\right)\nabla_\alpha \nabla_\beta R + \frac{1}{120} \nabla_\alpha \nabla_\beta R -\frac{1}{40} \square R_{\alpha\beta} +\frac{1}{30} R_{\alpha}\,^{\lambda} R_{\lambda\beta} -\frac{1}{60} {R^\rho\,_\alpha}^{\lambda}\,_{\beta} R_{\rho\lambda} -\frac{1}{60} R_{\lambda\mu\rho\alpha} R^{\lambda\mu\rho}{}_{\beta}.
\label{v10}
\end{eqnarray}
We shall restrict our calculation up to the quadratic order of the curvature.  We next collect all such terms to write a local momentum space representation of the Schwinger-DeWitt series expansion of the Feynman propagator,
\begin{eqnarray}
\bar{G}(k) =\frac{1}{k^2 +m^2} +\frac{\left(\frac16 -\xi\right)R}{(k^2 +m^2)^2} +\frac{i}{2} {\left(\frac16 -\xi\right)(\nabla_\alpha R}){\p^{\alpha}_k}(k^2 +m^2)^{-2} +\frac13 a_{\alpha\beta}\p^{\alpha}_k\p^{\beta}_k(k^2 +m^2)^{-2} \nonumber\\ +\left[\left(\frac16 -\xi\right)^2 {R^2} -\frac23 a^{\lambda}\,_{\lambda}\right]\frac{1}{(k^2 +m^2)^3} +{\cal O}(R^3).
\label{v11}
\end{eqnarray}
Note that the above propagator is locally Lorentz invariant, by the virtue of \ref{v4}. The above equation will serve as the key ingredient for our computations appearing below.

%%%%%%%%%%
\section{One loop effective potential at zero temperature}\label{S3} 
%%%%%%%%

The computation of the divergent part and subsequent renormalisation of the one loop effective action in a curved spacetime can be seen in, e.g.~\cite{Toms} and references therein. In this section, we wish to compute below the finite part of the effective potential at zero temperature after renormalisation, up to quadratic order in the curvature. Although the result appearing in this section can easily be found by directly using the standard `$\ln {\rm det}$' formula for the one loop effective action, we wish to briefly sketch below the use of the  local momentum space,  for the sake of completeness.  

We make the usual  decomposition for the field in \ref{v1} in terms of the classical background ($\phi$) and fluctuations ($\delta \phi$), $\Phi=\phi+\delta\phi$. Due to the background field the rest mass term appearing in \ref{v11} gets modified as,
 $m^2  \to m^2_1$ (say) where
\be
m^{2}_{1} = m^2 + \frac12 \lambda \phi^2 + \eta \phi.
\label{m1}
\ee
Clearly, for the discussion of SSB, the cubic self interaction is not relevant. However, we wish to keep the same in our general derivation for the effective potential, firstly because the $\lambda \phi^4+\eta \phi^3$ theory is a general renormalisable model in $(3+1)$-dimensions and second, the antisymmetric cubic self-interaction complements the symmetric and bounded nature of the quartic one. Most importantly, there can be various instances other than SSB where the asymmetry of the $\lambda \phi^4+\eta \phi^3$ self interaction could bring in some interesting physical effects.

\begin{figure}[H]
\begin{center}
\includegraphics[scale=0.22]{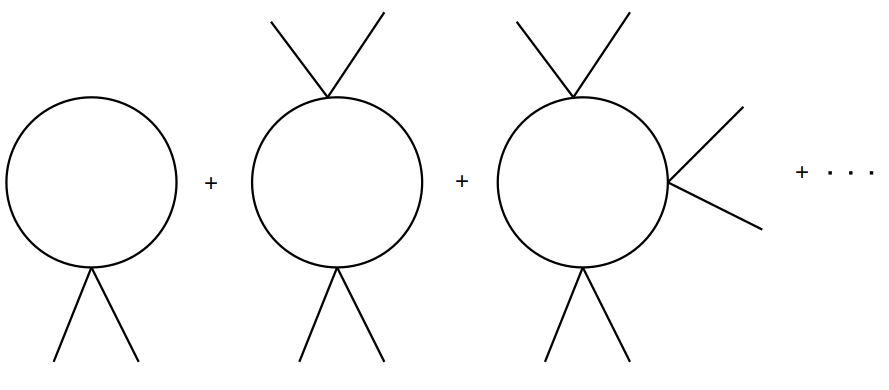}\\
\includegraphics[scale=0.22]{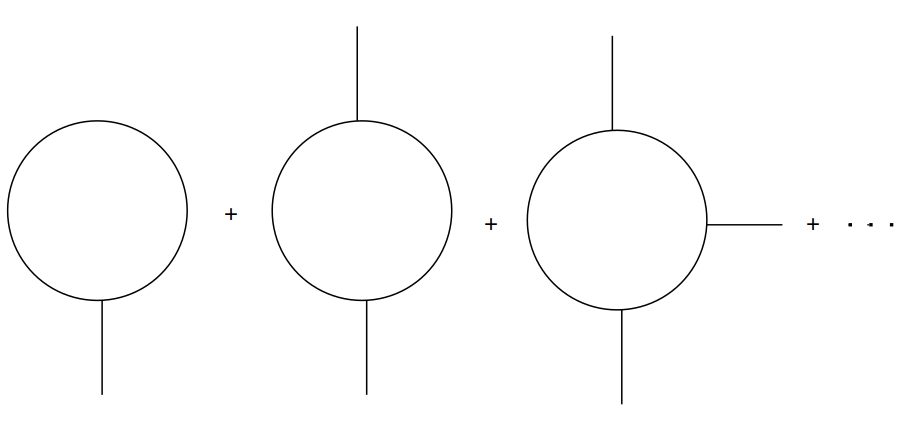}\\
\includegraphics[scale=0.32]{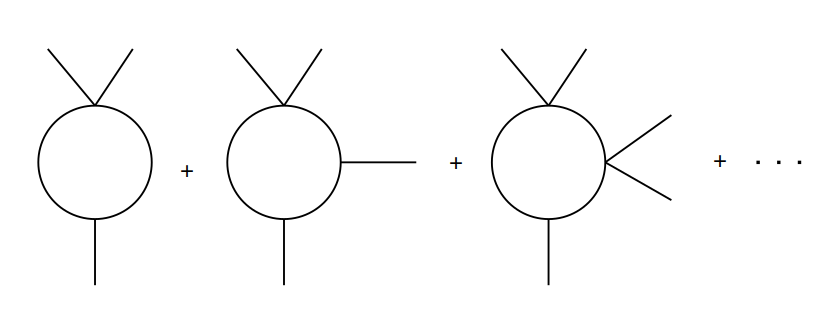}
\caption{\small \it The one loop diagrams for quartic plus cubic self interactions. The above three rows respectively contain quartic, cubic and mixed interaction vertices. Each row contains infinite number of one loop diagrams, found by inserting vertices on the vacuum loop. The vertices correspond to the background field, and no momenta flow across them.}
\label{fig1}
\end{center}
\end{figure}
\ref{fig1} represents the diagrammatic expansion of the one loop effective action. The same is basically the sum of all one loop diagrams found via vertex insertions shown in this figure. In the Minkowski spacetime, the one loop effective potential reads
\begin{eqnarray}
V_{\rm{eff}}^{1-{\rm loop}} = - \frac{\hbar}{2}\int \frac{d^dk}{(2\pi)^d} \ln G_{\rm{Min}}(k,m_1),
\label{v13}
\end{eqnarray}
where $G_{\rm{Min}}=(k^2+m_1^2)^{-1}$ is the  propagator of the scalar in the flat spacetime with an effective mass squared $m_1^2$, \ref{m1}. 
We now  replace the flat space propagator by that of the curved spacetime, \ref{v11},  to find
\begin{eqnarray}
&&V_{\rm{eff}}^{1-{\rm loop}} = -\frac{\hbar}{2}\int \frac{d^dk}{(2\pi)^d} \ln\left[\frac{1}{(k^2 +m^{2}_{1})} + \frac{(\frac16 -\xi)R}{(k^2 +m_{1}^{2})^2} + \frac{2f_2}{(k^2 +m_{1}^{2})^3}\right]  
\nonumber\\ &&= \frac{\hbar}{2}\int \frac{d^dk}{(2\pi)^d} \left[\ln (k^2 +m^{2}_{1}) - \frac{(\frac16 -\xi)R}{(k^2 +m_{1}^{2})} -\frac{2f_{2}}{(k^2 +m_{1}^{2})^2} +\frac12\frac{(\frac16 -\xi)^2 R^2}{(k^2 +m_{1}^{2})^2} \right] +{\cal O}(R^3),
\label{v15}
\end{eqnarray}
where the coefficient $f_2$ is quadratic in curvature, 
$$f_2 = \frac12\left(\frac16 -\xi\right)^2 {R^2} -\frac13 a^{\lambda}\,_{\lambda}$$
 and $a_{\mu\nu}$ is given by \ref{v10}.
The first  term on the right hand side of \ref{v15} is the flat space term, whereas the rest are the leading  curvature corrections. For $d=4-\e$, the three standard momentum integrals appearing above are given by  
\begin{eqnarray}
&&\int\frac{d^dk}{(2\pi)^d}\ln(k^2 +m_{1}^{2}) = \frac{m^{4}_{1}}{(4\pi)^2} \left[-\frac{\mu^{-\epsilon}}{\epsilon} +\frac12 \ln\frac{m_{1}^{2}}{4\pi\mu^2}- \frac14 -\frac12\psi(2) \right] +{\cal O}(\e),
\nonumber\\
&&\int\frac{d^dk}{(2\pi)^d}\frac{1}{(k^2 +m_{1}^{2})}=  \frac{ m^{2}_{1}}{(4\pi)^2} \left[-\frac{2 \mu^{-\epsilon}}{\epsilon} + \ln\frac{m_{1}^{2}}{4\pi\mu^2} -\psi(2) \right]+{\cal O}(\e), \nonumber\\
&&\int\frac{d^dk}{(2\pi)^d}\frac{1}{(k^2 +m_{1}^{2})^2} =\frac{1}{(4\pi)^2}\left[\frac{2\mu^{-\epsilon}}{\epsilon} +\psi(1) -\ln\frac{m^{2}_{1}}{4\pi\mu^2}\right]+{\cal O}(\e),
\label{v19}
\end{eqnarray}
where $\psi(z) =\Gamma^\prime (z)/\Gamma (z)$ is the digamma function and $\mu$ is a renormalisation scale. Substituting the three above integrals into \ref{v15} yields the (unrenormalised) one loop effective action.  The renormalisation of the effective potential or the action can be seen in~\cite{Toms}. The {\it purely divergent} part of the effective potential reads (with $\hbar=1$)
\begin{eqnarray}
&&V_{\rm{eff, \ div.}}^{1-{\rm loop}} = - \frac{ \mu^{-\e}}{16 \pi^2 \epsilon} \left[\left(\frac{1}{72} +\frac{\xi^2}{2}  -\frac{\xi}{6}  \right) R^2 +\frac{1}{90} \left(R_{\mu\nu\rho\sigma} R^{\mu\nu\rho\sigma} - R_{\mu\nu} R^{\mu\nu}\right) +\frac{m^4 }{2} +m^2 \eta \phi +\left(\xi -\frac{1}{6}  \right)m^2  R \right.\nonumber\\&&\left.+\left(\xi-\frac16 \right)\eta R \phi +\frac12 \left(m^2 \lambda +\eta^2 \right)\phi^2 +\frac{\lambda^2 \phi^4}{8}  +\frac{\eta\lambda\phi^{3}}{2}   +\frac{\lambda}{2}\left(\xi -\frac16  \right)R\phi^2 -\frac13\left(\xi -\frac15\right)\square R  \right].
\label{v27}
\end{eqnarray}
We may ignore the last term as it is a total divergence.  However note that, since \ref{v15} and \ref{v27} contain $\phi$-independent terms, probably one should not regard them as effective potential. In a strict sense,  they should instead be attributed as effective Lagrangian density minus the kinetic term.

In order to renormalise the divergence associated with the pure curvature terms of \ref{v27}, we take the gravitational counterterms in the Lagrangian density
\begin{eqnarray}
\delta{\cal L}_{\rm grav} = \left[\delta \Lambda +\delta \kappa R +\delta \alpha_{1} R_{\mu\nu\rho\sigma} R^{\mu\nu\rho\sigma} +\delta \alpha_{2} R_{\mu\nu} R^{\mu\nu} +\delta \alpha_{3} R^2 \right],
\label{v25}
\end{eqnarray}
with the choice
\begin{eqnarray}
\delta \Lambda = \frac{\mu^{-\e}m^4}{32 \pi^2 \epsilon},  \qquad \delta\kappa =\frac{\mu^{-\e}m^2\left(\xi  -\frac16 \right)}{16 \pi^2 \epsilon}, \qquad
\delta\alpha_{1} =\frac{\mu^{-\e}}{16 \pi^2 \epsilon}\frac{1}{90},\nonumber\\
\delta\alpha_{2} =-\frac{\mu^{-\e}}{16 \pi^2 \epsilon}\frac{1}{90}, \qquad
\delta\alpha_{3} =\frac{\mu^{-\e}}{32 \pi^2 \epsilon}\left(\frac{1}{36} + \xi^2 -\frac{\xi}{3} \right).\qquad 
\label{v25'}
\end{eqnarray}
The remaining divergence of \ref{v27} is absorbed by a matter counterterm Lagrangian 
\begin{eqnarray}
\delta{\cal L}_{\rm matt}  = \left[\frac12 \delta m^{2}\phi^2  +\frac12 \delta\xi R \phi^2  +\frac{\delta \eta \phi^{3}}{3!} +\frac{\delta \lambda \phi^{4}}{4!} +\delta \tau \phi +\delta \gamma R \phi \right],
\label{v24}
\end{eqnarray}
with the choice
\begin{eqnarray}
&&\delta m^{2} =-\frac{\mu^{-\e}\left(m^2 \lambda +\eta^2 \right)}{16 \pi^2 \epsilon}, \qquad
\delta\xi =- \frac{\mu^{-\e}\lambda\left(\xi -\frac16  \right)}{16 \pi^2 \epsilon},\qquad
\delta\tau = -\frac{\mu^{-\e} m^2\eta}{16 \pi^2 \epsilon},  \nonumber\\&&
\delta\gamma =- \frac{\mu^{-\e} \eta \left(\xi -\frac16 \right)}{16 \pi^2 \epsilon}, \qquad 
\delta\eta =- \frac{3\mu^{-\e}\eta\lambda}{16 \pi^2 \epsilon},  \qquad
\delta\lambda =- \frac{3\mu^{-\e} \lambda^2}{16 \pi^2 \epsilon}.
\label{v29}
\end{eqnarray}

 We further introduce  counterterms to absorb terms which are {\it independent} of the background field $\phi$, but otherwise {\it finite},
\begin{eqnarray}
&&\delta \Lambda_{\rm fin.} = \frac{m^4}{64\pi^2} \left(\psi(2)+\frac12 +\ln\frac{4\pi\mu^2}{m^{2}}\right),  \qquad \delta\kappa_{\rm fin.} =\frac{m^2(\xi  -1/6)}{32\pi^2}\left(\psi(2)+\ln\frac{4\pi\mu^2}{m^{2}}\right), \nonumber\\&&
\delta\alpha_{1, {\rm fin.}} =\frac{1}{2880\pi^2} \left(\psi(1)+\ln\frac{4\pi\mu^2}{m^{2}}\right),\qquad 
\delta\alpha_{2, {\rm fin.}} =-\frac{1}{{2880 \pi^2}}\left({\psi(1)}+\ln\frac{4\pi\mu^2}{m^{2}}\right), \nonumber\\&&
\delta\alpha_{3,{\rm fin.}} =  \frac{(\xi-1/6)^2}{{64 \pi^2}}\left({\psi(1)}+\ln\frac{4\pi\mu^2}{m^{2}}\right).
\label{v25''}
\end{eqnarray}
We will not absorb any field dependent finite terms by any counterterms, and will leave them intact. We thus finally have the renormalised result
\begin{eqnarray}
&&V_{\rm{eff, Ren}}^{1-{\rm loop}} = \frac12(m^2+\xi R)\phi^2+\frac{\lambda \phi^4}{4!}+\frac{\eta \phi^3}{3!} +\frac{1}{2 (4\pi)^2} \left[\frac{m^{4}_{1}}{2}\left( \ln\frac{m_{1}^{2}}{4\pi\mu^2}- \frac12 -\psi(2) \right) + \frac{m^4}{2} \left(\frac12+\psi(2) \right)-\frac{m^4}{2}\ln\frac{m^{2}}{4\pi\mu^2}\right.\nonumber\\&&\left. +\left(\frac16 -\xi\right)R \left(\psi(2)(m^{2}_{1}-m^2)-m_1^2 \ln\frac{m_{1}^{2}}{4\pi\mu^2}+m^2 \ln\frac{m^{2}}{4\pi\mu^2}\right)-\left(\frac12\left(\frac16 -\xi\right)^2 R^2 -2f_1\right) \ln \frac{m_1^2}{m^2} \right],
\label{v30'}
\end{eqnarray}
where $m_{1}^{2}$ is given by \ref{m1}, and we have abbreviated
$$f_1 = \frac12\left(\frac16 -\xi\right)^2 R^2 -\frac13 a^{\lambda}\,_{\lambda} = \frac12\left(\frac16 -\xi\right)^2 {R^2} +\frac{1}{180}R_{\mu\nu\rho\sigma} R^{\mu\nu\rho\sigma} -\frac{1}{180} R_{\mu\nu} R^{\mu\nu}.$$
Note that the renormalised result \ref{v30'} does not contain any term independent of the background field $\phi$, and becomes vanishing at $\phi=0$. Thus we may regard it as a true effective potential for the scalar, containing the curvature effects.

Let us now take the flat spacetime limit of \ref{v30'}, for a massless and quartic self interaction. Taking $\psi(2)\simeq  0.422785$ and setting $R=f_1=\eta=m=0$,   we reproduce the Coleman-Weinberg effective potential~\cite{Coleman}
\begin{eqnarray}
 V_{\rm{eff}}^{\rm flat} %&&= \frac{\lambda\phi^{4}}{4!}+\frac{\lambda^{2} \phi^{4}}{256 \pi^{2}}\left[\ln\frac{\lambda \phi^{2}}{\mu^{2}}- \ln(8\pi) -\frac12 -0.422785\right]
%\nonumber\\&&
 = \frac{\lambda\phi^{4}}{4!}+\frac{ \lambda^{2} \phi^{4}}{256 \pi^{2}}\left[\ln\frac{\lambda \phi^{2}}{\mu^{2}}-4.146\right].
\label{v32}
\end{eqnarray}

Finally, we  note that the derivation of the  counterterms of \ref{v29} is completely equivalent to that of the imposition of renormalisation conditions. For example, for the  mass renormalisation condition in curved spacetime we impose from \ref{v27}, \ref{v24}  and  the tree level potential,
  $$\left(V_{\rm{eff, \ div.}}^{1-{\rm loop}} +\delta{\cal L}_{\rm matt}+ \cal{L}_{\rm tree} \right)''_{\phi = 0} = (m^2+\xi R)$$
The above equation gives, 
  $$ \frac{\mu^{-\e}}{16\pi^2\e}\left[ (m^2\lambda+\eta^2) +\lambda\left(\xi-\frac16 \right)R\right]+ (m^2+\delta m^2)+(\xi+\delta \xi) R   = m^2+\xi R,$$
  which immediately yields expressions same as that of \ref{v29} for $\delta m^2$ and $\delta \xi$. The other field counterterms can be derived in a similar manner. We further refer our reader to \cite{Arai:2012sh, LopezNacir:2013alw} for a discussion on the equivalence of these two approaches in the context of $O(N)$ symmetric scalar field theory.

%%%%%%%%%%%
\section{The case  at finite temparature}\label{S4}\
%%%%%%%%%%%%
\noindent
We next wish to generalise  \ref{v15} for  finite temperature, $T=\beta^{-1}$. In order to obtain the pure thermal contribution, we recall that one needs replacing the Euclidean momentum $k_0$ by discrete $\omega_n$ where $\omega_n =2\pi n/\beta$ for bosons, i.e.~\cite{Guang, Hu}
\be
k^2 +m_{1}^{2} \to \vec{k}^2+ \left(\frac{2\pi n}{\beta}\right)^2 +m_{1}^{2}, \qquad  \qquad n=0, \pm 1, \pm 2, \pm 3, \cdots
\label{v34'}
\ee
This leads to the replacements in \ref{v15}
\begin{eqnarray}
\int\frac{d^dk}{(2\pi)^d}\ln(k^2 +m_{1}^{2}) \longrightarrow\sum_{n =-\infty}^{\infty}\frac{1}{\beta}\int\frac{d^{d-1}\vec{k}}{(2\pi)^{d-1}}\ln\left[\vec{k}^2+ \left(\frac{2\pi n}{\beta}\right)^2 +m_{1}^{2}\right],\nonumber\\
\int\frac{d^dk}{(2\pi)^d}\frac{1}{(k^2 +m_{1}^{2})^s} \longrightarrow\sum_{n=-\infty}^{\infty}\frac{1}{\beta}\int\frac{d^{d-1}\vec{k}}{(2\pi)^{d-1}}\frac{1}{\left[\vec{k}^2+ \left(\frac{2\pi n}{\beta}\right)^2 +m_{1}^{2}\right]^s} .
\label{v35}
\end{eqnarray}

Before we proceed for the effective action, let us see the effect of the curvature on the self energy in a thermal field theory, in order to further motivate the problem we are interested in. For a scalar field with mass $m$ and quartic self interaction, the renormalised one loop self energy reads~\cite{kapusta}
\begin{eqnarray}
\Sigma = 12\lambda\int \frac{d^3\vec k}{(2\pi)^3} \frac{1}{\left(\vec{k}^2+m^2\right)^{\frac12}}\frac{1}{e^{\beta\sqrt{\vec{k}^2+m^2}} -1}.
\label{g1}
\end{eqnarray}
Assuming the scalar to be light, the first two leading order terms read
\begin{eqnarray}
\Sigma = \lambda T^2 \left(1-\frac{3m}{\pi T} \right).
\label{g2}
\end{eqnarray}
This shows generation of a scalar mass due to the finite temperature. Let us now `turn on' the gravity  and for simplicity assume the spacetime to be weakly curved. Keeping only up to the ${\cal O}(R)$ term, we have from \ref{v11} 
\begin{eqnarray}
{\bar G}(k)= \left[1-\left(\frac16 -\xi \right)R \frac{\p}{\p m^2} \right]\frac{1}{k^2+m^2}.
\label{g3}
\end{eqnarray}
Promoting now the propagator to thermal case as stated above, we have the renormalised one loop self energy 
\begin{eqnarray}
\Sigma = \lambda T^2 -\frac{3\lambda mT}{\pi}-\left(\frac16 -\xi \right)R\frac{\p}{\p m^2}\left(\lambda T^2 -\frac{3\lambda mT}{\pi}\right)=\lambda T^2 -\frac{3\lambda mT}{\pi}+\left(\frac16 -\xi \right)\frac{3\lambda R T}{2\pi m}.
\label{g4}
\end{eqnarray}
The third term of the final result exhibits mass generation in a thermal field theory in a weakly curved spacetime.  We  note that if the scalar is extremely light, the curvature correction term becomes very large, making the result untrustworthy. It seems that one needs to develop some non-perturbative resummation technique in order to tackle such an issue.   Nevertheless, the above result indeed indicates that the spacetime curvature can have interesting effects   on thermal fields in a curved spacetime. \\

\noindent
Let us now come to the computation of the effective potential. In \ref{v35}, we may break the summation by separating the zero mode as 
\begin{eqnarray}
\sum_{n =-\infty}^{\infty} \equiv (n=0)+ 2\sum_{n =1}^{\infty}.
\label{v36}
\end{eqnarray}
We shall essentially work in the high temperature approximation $(m \beta \ll 1)$ and hence will make a power series expansion in $\beta$ for relevant quantities, and will retain terms up to ${\cal O}(\beta^6)$ in the following.

Now, the pure thermal part of the  first integral of \ref{v19} becomes at finite temperature
\begin{eqnarray}
\sum_{n =1}^{\infty}\frac{1}{\beta}\int\frac{d^{d-1}\vec{k}}{(2\pi)^{d-1}}\ln\left[\vec{k}^2+ \left(\frac{2\pi n}{\beta}\right)^2 +m_{1}^{2}\right] =\sum_{n =1}^{\infty}\frac{1}{\beta}\frac{2}{(d-1) (4\pi)^{\frac{d-1}{2}}} \left[\left(\frac{2\pi n}{\beta}\right)^2 +m_{1}^{2}\right]^{\frac{d-1}{2}}\Gamma\left(\frac{3-d}{2}\right) .
\label{v37}
\end{eqnarray}
We have,
\begin{eqnarray}
&&\sum_{n =1}^{\infty}\left[\left(\frac{2\pi n}{\beta}\right)^2 +m_{1}^{2}\right]^{\frac{d-1}{2}}  =\sum_{n =1}^{\infty}m_{1}^{d-1}\left(n\alpha\right)^{d-1}\left[1+\frac{1}{(n\alpha)^2}\right]^{\frac{d-1}{2}}, 
\label{v38}
\end{eqnarray}
where we have defined  $\alpha= 2\pi/(m_1 \beta)$.
Expanding for a small $\beta$, the above expression equals
\begin{eqnarray}
 m_{1}^{d-1}\alpha^{-\epsilon}\left[\alpha^3 \zeta(-3)+\frac32 \alpha \zeta(-1)+\frac38 \alpha^{-1}\zeta(1+\epsilon)- \frac{1}{16}\alpha^{-3}\zeta(3)+\frac{1}{128}\alpha^{-5}\zeta(5)-\frac{5}{3072}\alpha^{-7}\zeta(3) + {\cal O}(\beta^8)\right],
\label{v39}
\end{eqnarray}
where $\zeta(n) =\sum_{j=1}^{\infty}j^{-n}$ is the Riemann zeta function. Using the properties of this function pertaining to its regularisation,  \ref{v37} becomes
\begin{eqnarray}
\sum_{n =1}^{\infty}\frac{1}{\beta}\int\frac{d^{d-1}\vec{k}}{(2\pi)^{d-1}}\ln\left[\vec{k}^2+ \left(\frac{2\pi n}{\beta}\right)^2 +m_{1}^{2}\right] = -\frac{1}{90}\frac{\pi^2}{\beta^4}+\frac{1}{24}\frac{m_{1}^{2}}{\beta^2}-\frac{m_{1}^{3}}{12\pi \beta}-\frac{m^{4}_{1} \ln{m^{2}_{1}\beta^2}}{64\pi^2} +\frac{m^{4}_{1}\left(\psi(1) +\ln{4\pi^2}\right)}{64\pi^2} \nonumber\\ +\frac{m^{6}_{1}\beta^2\zeta(3)}{768\pi^4}-\frac{m^{8}_{1}\beta^4\zeta(5)}{24576\pi^6}+\frac{m^{10}_{1}\beta^6\zeta(7)}{196608\pi^8}.
\label{v42}
\end{eqnarray}
Likewise we have 
\begin{eqnarray}
&&-\sum_{n =1}^{\infty}\left(\frac16 -\xi\right)R \int\frac{d^{d-1}\vec{k}}{(2\pi)^{d-1}}\frac{1}{\left[\vec{k}^2+ \left(\frac{2\pi n}{\beta}\right)^2 +m_{1}^{2}\right]} =-\left(\frac16 -\xi\right)R\left[\frac{1}{24\beta^2}-\frac{m_1}{8\pi \beta}+\frac{m_{1}^{2}}{16\pi^2}\left(\psi(1) +\ln{\frac{2\pi}{m_{1} \beta}}\right)\right. \nonumber\\&& \left. +\frac{m_{1}^{4} \beta^2}{256\pi^4}\zeta(3) - \frac{m_{1}^{6} \beta^4}{6144\pi^6}\zeta(5)+ \frac{5m_{1}^{8} \beta^6}{196608\pi^8}\zeta(7) \right]
\label{v43}
\end{eqnarray}
and
\begin{eqnarray}
&&\sum_{n =1}^{\infty}\left(\frac12 \left(\frac16 -\xi\right)^2 R^2 -2f_1\right)\int\frac{d^{d-1}\vec{k}}{(2\pi)^{d-1}}\frac{1}{\left[\vec{k}^2+ \left(\frac{2\pi n}{\beta}\right)^2 +m_{1}^{2}\right]^2} = \left(\frac12 \left(\frac16 -\xi\right)^2 R^2 -2f_1\right) \left[\frac{1}{16\pi\beta m_1}\right. \nonumber\\&& \left. -\frac{1}{16\pi^2}\left(\psi(1) +\ln{\frac{2\pi}{m_{1} \beta}}\right)   - \frac{m_{1}^{2} \beta^2}{128\pi^4}\zeta(3)+\frac{m_{1}^{4} \beta^4}{2048\pi^6}\zeta(5)-\frac{5 m_{1}^{6} \beta^6}{49152\pi^8}\zeta(7) \right].
\label{v45}
\end{eqnarray}
An alternative and more rigorous derivation of the above integrals in the flat spacetime by breaking the Feynman propagator into temperature independent and dependent parts can be seen in~\cite{dolan}. We also note that in the three integrals appearing above there are a couple of temperature independent parts, such as the fifth term appearing on the right hand side of \ref{v42}, even though  
\ref{v42}, \ref{v43}, \ref{v45} are pure thermal contributions. Such $\beta$-independent parts originate from the regularisation of the $\zeta$-function. 

Putting things together now, we obtain the total renormalised one loop effective potential at the high temperature limit
\begin{eqnarray}
&&V_{\rm{eff},\beta}^{1-{\rm loop}} (\beta^{-1} \gg 1)= \frac12 (m^2+\xi R)\phi^2 + \frac{\lambda \phi^4}{4!}+\frac{\beta \phi^3}{3!} + V_{\rm{eff}}^{1-{\rm loop}} (\beta^{-1}=0)+\left[-\frac{\pi^2}{90\beta^4}+\frac{m_{1}^{2}}{24\beta^2}-\frac{m_{1}^{3}}{12\pi \beta}\right. \nonumber\\&& \left. -\frac{ m^{4}_{1} \ln{m^{2}_{1}\beta^2}}{64\pi^2}+\frac{m^{4}_{1}\left(\psi(1) +\ln{4\pi}\right)}{64\pi^2}   +\frac{m^{6}_{1}\beta^2\zeta(3)}{768\pi^4}-\frac{m^{8}_{1}\beta^4\zeta(5)}{24576\pi^6}+\frac{m^{10}_{1}\beta^6\zeta(7)}{196608\pi^8} \right. \nonumber\\&& \left. -\left(\frac16 -\xi\right)R\left(\frac{1}{24\beta^2}-\frac{m_1}{8\pi \beta}+\frac{m_{1}^{2}}{16\pi^2}\left(\psi(1) +\ln\frac{2\pi}{m_{1} \beta}\right) +\frac{m_{1}^{4} \beta^2}{256\pi^4}\zeta(3) - \frac{m_{1}^{6} \beta^4}{6144\pi^6}\zeta(5) + \frac{5m_{1}^{8} \beta^6}{196608\pi^8}\zeta(7)\right) \right. \nonumber\\&& \left. +\left( \frac12 \left(\frac16 -\xi\right)^2 R^2-2f_1\right) \left(\frac{1}{16\pi\beta m_1}-\frac{1}{16\pi^2}\left(\psi(1) +\ln\frac{2\pi}{m_{1} \beta}\right) - \frac{m_{1}^{2} \beta^2}{128\pi^4}\zeta(3) +\frac{m_{1}^{4} \beta^4}{2048\pi^6}\zeta(5)-\frac{5 m_{1}^{6} \beta^6}{49152\pi^8}\zeta(7) \right)              +{\cal O}(\beta^7)\right], \nonumber\\
\label{v46}
\end{eqnarray}
where $m_1^2= m^2+ \lambda \phi^2/2+\eta \phi$, as earlier and $V_{\rm{eff}}^{1-{\rm loop}}(\beta^{-1}=0)$ is the zero temperature effective potential given by \ref{v30'}. Note that there is no temperature dependent divergence present in the above expression. This shows that renormalisation counterterms are independent of $\beta$, as is expected~\cite{kapusta}. Also, while investigating the behaviour of the above thermal effective potential, we will ignore all terms  independent of $\phi$ but otherwise finite.

As we have mentioned  earlier, the above one loop effective potential was first derived in~\cite{Hu} for $\eta=0$. However, our notation is different compared to this work. We also note from~\cite{Hu} that   
if we set the arbitrary mass scale $\mu=\beta^{-1}$ in \ref{v30'}, {\it all} the logarithmic terms appearing in the finite temperature part of the effective potential precisely cancel with that of the zero temperature. However, we wish to keep $\mu$ to be a free parameter in this work. Such terms will contribute non-trivially to the feature of the effective potential and hence to the spontaneous symmetry breaking phenomenon. \\

\noindent
Although we are interested in the high temperature limit of the effective action, we wish to briefly discuss the low temperature limit of the same below. For $\beta^{-1}\to 0$, we make the expansion
\begin{eqnarray}
&&\left[\left(\frac{2\pi n}{\beta}\right)^2 +m_{1}^{2}\right]^{\frac{d-1}{2}}  = m_{1}^{3}\left[1+\frac32 \left(\frac{2\pi n}{m_{1} \beta}\right)^2+\frac38 \left(\frac{2\pi n}{m_{1}^{2} \beta}\right)^4+ \cdots \right].
\label{v47} 
\end{eqnarray}
Putting this expression back into \ref{v37}, we obtain
\begin{eqnarray}
&&\sum_{n =1}^{\infty}\frac{1}{\beta}\int\frac{d^{d-1}\vec{k}}{(2\pi)^{d-1}}\ln\left[\vec{k}^2+ \left(\frac{2\pi n}{\beta}\right)^2 +m_{1}^{2}\right]_{\beta^{-1}\to 0}\approx \sum_{n =1}^{\infty}\frac{1}{\beta}\frac{2 m_{1}^{3}}{(d-1) (4\pi)^{\frac{d-1}{2}}} \left(1+\frac32 \left(\frac{2\pi n}{m_{1} \beta}\right)^2+\frac38 \left(\frac{2\pi n}{m_{1} \beta}\right)^4\right)\Gamma\left(\frac{3-d}{2}\right). \nonumber\\
\label{v48} 
\end{eqnarray}
The other relevant integrals can be approximated in a likewise manner. Using now Ramanujan's result, $\zeta{(-2k)}=0$ ($k=\ {\rm integer}$), we obtain after a little bit of algebra
\begin{eqnarray}
&&V_{\rm{eff},\beta^{-1}\ll 1}=\frac12 m^{2} \phi^2 + \frac{\eta \phi^3}{3!}  + \frac{\lambda\phi^4}{4!}+\frac{\xi R\phi^2}{2}+V_{\rm{eff}}(\beta^{-1}=0) \nonumber\\
&& +\frac32\left[- \frac{m_{1}^{3}}{3\pi\beta}+\left(\frac16-\xi \right)\frac{m_1 R}{2\pi\beta} +\left(-f_1+\frac14\left(\frac16-\xi \right)^2R^2\right)\frac{1}{\pi m_1\beta} \right].
\label{v51}
\end{eqnarray}
Setting $\beta^{-1} \to 0$ is possible in \ref{v51} whereas it is not possible in \ref{v46}. 

We further wish to consider two loop computations for the effective potential. However, before we do that, let us discuss the issue of spontaneous symmetry breaking and phase transition with the one loop high temperature effective potential,  \ref{v46}, we have found. We also note in \ref{v46} that for $\eta \neq 0$, and since both $\eta$ and $\phi$ can be positive or negative, the expression becomes invalid for $m_1^2\to 0$, owing to the $\ln m_1^2$ term. In other words, the effective potential will be valid in the parameter or the field space region where $m_1^2 \ensuremath{>} 0$. Owing to the fact that the background field $\phi$  is real, it is easy to see that this condition implies, 
\be
\left(\frac{\eta}{m}\right)^2 \ensuremath{<} 2\lambda .
\label{add}
\ee
 We shall keep this in mind in the following computations 
 whenever relevant.

%%%%%%%%%%
\section{Spontaneous symmetry breaking and phase transition at zero temperature}\label{S6}
%%%%%%%%%%%%

The chief motivation of this section is to see the effect of the spacetime curvature in the phase transition or spontaneous symmetry breaking phenomenon, at zero temperature. As we have already mentioned, this topic is old. In flat spacetime, one of the earliest thermal field theory calculations in this context can be seen in the seminal work~\cite{dolan}. At zero temperature and with spacetime curvature some such computations can be seen in e.g.~\cite{Toms, Ishi, Buch} and references therein. In particular, \cite{Ishi, Buch} use weak or  linear in curvature approximation and renormalisation group techniques. However, as we have also mentioned earlier, such results cannot be applicable to Ricci flat spacetimes.

\begin{figure}[H]
\begin{center}
\includegraphics[scale=0.4]{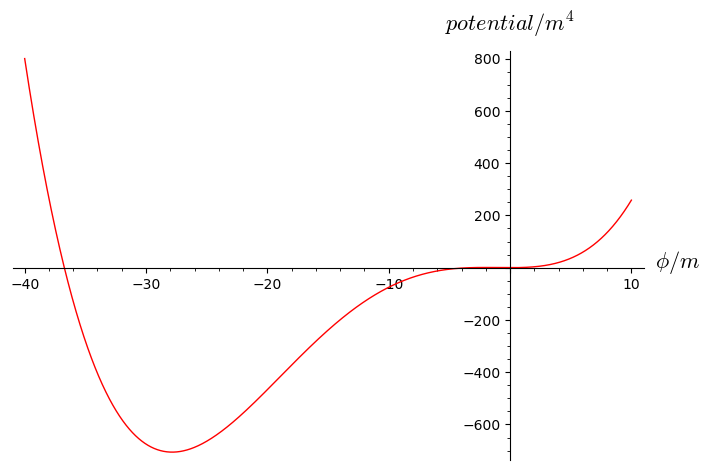}
\caption{\it \small Plot of the tree level dimensionless  self interaction potential, $ V(\bar{\phi})= \bar{\eta}\bar{\phi}^3/3! + \lambda\bar{\phi}^4/4!$. We have taken $\lambda = 0.1$ and $\bar{\eta}=0.447$.}
\label{f1add}
\end{center}
\end{figure}

Let us now make the field and the cubic coupling constant dimensionless by scaling them with respect to its rest mass, which we assume to be non-vanishing, i.e., $\bar{\phi}= \phi/m$ and $\bar{\eta}=\eta/m$. The tree level asymmetric self interaction potential has been plotted in  \ref{f1add}.

\begin{figure}[H]
\begin{center}
\end{center}
\centering
\begin{subfigure}{.4\textwidth}
  \centering
  \includegraphics[scale=0.45]{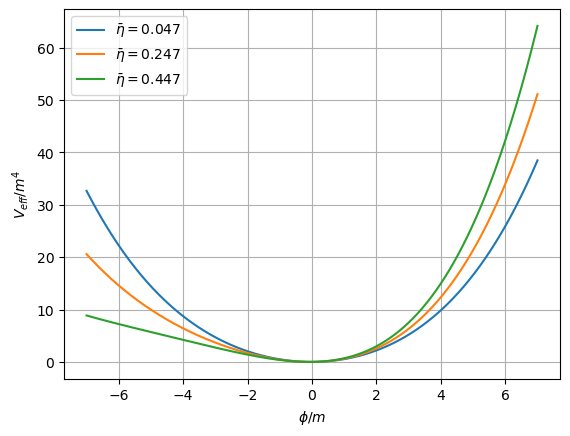}
  \caption{$\eta\phi^3 +\lambda\phi^4$ interaction, $\lambda$=0.1}
  \label{fig:sub3}
\end{subfigure}%
\begin{subfigure}{.7\textwidth}
  \centering
  \includegraphics[scale=0.45]{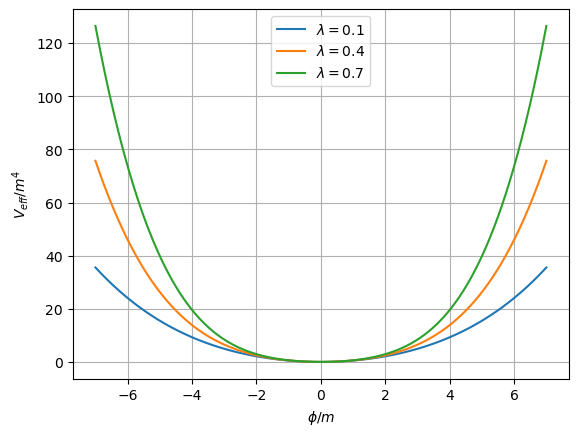}
  \caption{$\lambda\phi^4$ interaction}
  \label{fig:sub4}
\end{subfigure}
\caption{\it \small The one loop effective potential, \ref{v54}, with vanishing temperature in flat spacetime. Note that for $\bar{\eta}=0$, no spontaneous symmetry breaking is indicated, as we have taken  the rest mass of the  scalar to be non-vanishing.  Note that by our renormalisation scheme, the effective potential does not contain any field independent finite terms.}  
\label{fig:test2}
\end{figure}

Let us now consider the zero temperature effective potential in the absence of curvature, \ref{v30'} ($R=0=f_1$), in the dimensionless form
\begin{eqnarray}
&&\frac{V_{\rm{eff}}}{m^4} = \frac{\bar{\phi}^2}{2}   + \frac{\bar{\eta}\bar{\phi}^3}{3!}  + \frac{\lambda\bar{\phi}^4}{4!} +\frac{1}{4 (4\pi)^2} \left[\left(2\left(\frac12\lambda \bar{\phi}^2 + \bar{\eta}\bar{\phi} \right)+\left(\frac12\lambda \bar{\phi}^2+\bar{\eta}\bar{\phi}  \right)^2\right) \left(\ln \frac{1+\frac12\lambda\bar{\phi}^2+\bar{\eta}\bar{\phi} }{\frac{4\pi \mu^2}{m^2}}-\frac12-\psi(2)\right) \right. \nonumber\\&& \left.+ \ln \left(1+\frac12\lambda \bar{\phi}^2+\bar{\eta}\bar{\phi}  \right) \right].
\label{v54}
\end{eqnarray}
Putting in now some customary numerical values, $m \sim  0.0005 {\rm GeV}$ and  $\mu\sim  3\times10^{-25} {\rm GeV} $, which may be appropriate for the large scale physics we are interested in, e.g.~\cite{Martin} and references therein, we have plotted the above zero temperature flat space effective potential in \ref{fig:test2}. We note that since the scalar we have taken has non-vanishing rest mass, there can be no spontaneous symmetry breaking for the quartic self interaction. However, in the presence of spacetime curvature, the scenario can change, as we wish to show below. 

\noindent
We wish to consider two physically well motivated spacetimes below, namely, the Schwarzschild and the de Sitter spacetime.\\

\noindent
{\underline{\bf The Schwarzschild spacetime}}:\\

\begin{figure}[H]
\begin{center}
\end{center}
\centering
\begin{subfigure}{.4\textwidth}
  \centering
  \includegraphics[scale=0.45]{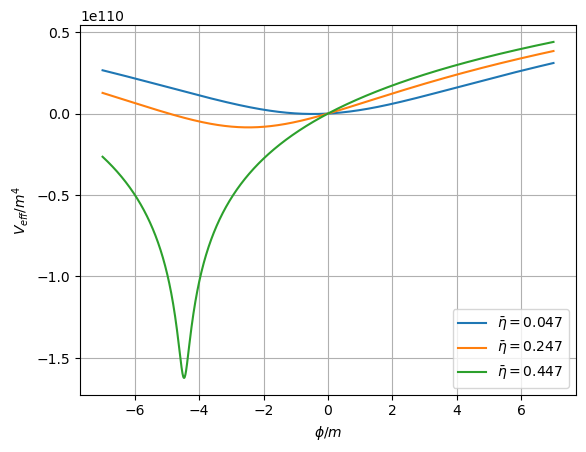}
  \caption{$\eta\phi^3 +\lambda\phi^4$ interaction, $\lambda=0.1$}
  \label{fig:sub5}
\end{subfigure}%
\begin{subfigure}{.7\textwidth}
  \centering
  \includegraphics[scale=0.45]{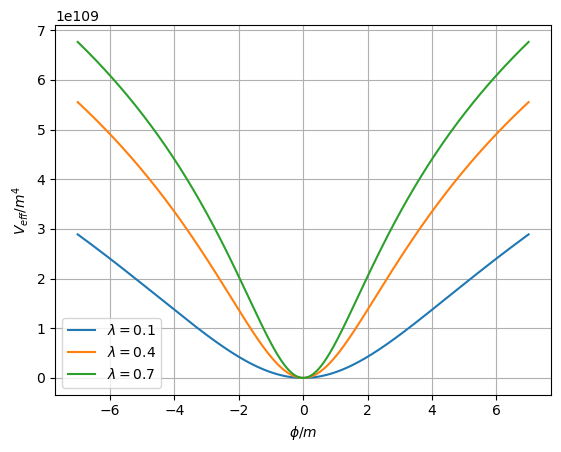}
  \caption{$\lambda\phi^4$ interaction}
  \label{fig:sub6}
\end{subfigure}
\caption{{\it \small The one loop effective potential, \ref{v55}, at vanishing temperature in the Schwarzschild spacetime. See main text for discussion. } }
\label{fig:test3}
\end{figure}

\noindent
For the Schwarzschild spacetime, 
$$ds^2 = \left( 1- \frac{2MG}{r}\right)dt^2 + \left( 1- \frac{2MG}{r}\right)^{-1}dr^2 + r^2 d\Omega^2,$$
we must set $R=0=R_{\mu\nu}$, and 
$$R^{\mu\nu\rho\sigma}R_{\mu\nu\rho\sigma}=\frac{48G^2M^2}{r^6}, $$
in \ref{v30'}, so that we have around $r \sim 2GM$
\begin{eqnarray}
&&\frac{V_{\rm{eff}}}{m^4} = \frac{\bar{\phi}^2}{2}  + \frac{\bar{\eta}\bar{\phi}^3}{3!}  + \frac{\lambda\bar{\phi}^4}{4!} +\frac{1}{4 (4\pi)^2} \left[\left(2\left(\frac12\lambda \bar{\phi}^2 + \bar{\eta}\bar{\phi} \right)+\left(\frac12\lambda \bar{\phi}^2+\bar{\eta}\bar{\phi}  \right)^2 \right)\left(\ln \frac{1+\frac12\lambda\bar{\phi}^2+\bar{\eta}\bar{\phi}  }{\frac{4\pi \mu^2}{m^2}}-\frac12-\psi(2)\right) \right. \nonumber\\&& \left.+ \ln \left(1+\frac12\lambda \bar{\phi}^2+\bar{\eta}\bar{\phi}   \right) \right]+\frac{0.0016}{m^4 G^4 M^4} \ln \left(1+\frac12\lambda\bar{\phi}^2+\bar{\eta}\bar{\phi} \right). 
\label{v55}
\end{eqnarray}
 Note that the effective potential diverges as $M \to 0$. This corresponds to the fact that the value of $R^{\mu\nu\rho\sigma}R_{\mu\nu\rho\sigma}$ increases on the horizon with decreasing mass of the black hole. On the other hand, the result approaches the flat spacetime limit, \ref{v54}, for large value of the black hole mass.   Taking now for example, $GM \sim 10^2{\rm m } \sim 10^{17}{\rm GeV}^{-1}$,
we have plotted \ref{v55} in \ref{fig:test3}. We note in particular for the first of \ref{fig:test3}, that for increasing cubic coupling, the effective potential develops a deep. This  originates from the logarithm contained in the last term on the right hand side of \ref{v55}, which may contribute a large negative value. Note also that for a given $\lambda$ value, we cannot arbitrarily    go on increasing the cubic coupling, owing to the discussion appearing at the end of the preceding Section, \ref{add}.

\noindent
Let us also consider  in this context the Kerr spacetime,
\begin{eqnarray}
&&ds^2 = -\frac{r^2-2GMr+a^2}{r^2+a^2 \cos^2\theta}dt^2-\frac{4aMGr \sin^2\theta}{r^2+a^2 \cos^2\theta}dt d\phi +\frac{r^2+a^2 \cos^2\theta}{r^2-2MGr+a^2}dr^2+\left(r^2+a^2 \cos^2\theta\right)d\theta^2 \nonumber\\&&+\left(r^2+a^2+\frac{2a^2GMr \sin^2\theta}{r^2+a^2 \cos^2\theta}\right)\sin^2\theta d\phi^2 ,
\label{v55'}
\end{eqnarray}
which is also Ricci flat, and the Kretschmann scalar reads
$$R_{\mu\nu\lambda \rho}R^{\mu\nu\lambda \rho}=\frac{48G^2M^2\left(r^6-15a^2r^4\cos^2\theta+15a^4r^2\cos^2\theta-a^6\cos^6\theta \right)}{\left(r^2+a^2 \cos^2\theta\right)^6}.$$
Comparing this with the Schwarzschild spacetime, we see that there is no qualitative change at all due to spacetime rotation on the equatorial plane, $\theta=\pi/2$. The effect of the spacetime rotation is maximum on the axis ($\theta=0,\pi$). We note that if we take the extremal limit ($a\to GM$), the Kretschmann scalar vanishes as we move towards  the horizon ($r\to GM$) along the axis. This shows that the one loop effective potential for an extreme Kerr black hole spacetime  in the vicinity of the axial points near the horizon coincides with that of the flat spacetime,~\ref{v54}.

 Finally, we note that neither the Schwarzschild nor the Kerr case shows any SSB. This is because, as can be  easily checked, that \ref{v55} is always  a positive definite for $\bar \eta=0$, and the effective potential always must pass through the origin, for all values of the masses $m$ and $M$. Second, changing these mass parameters does not create any new extrema.  However, the situation will be qualitatively different for the de Sitter spacetime, as follows. \\

\noindent
{\underline{\bf The de Sitter spacetime}}:\\

\begin{figure}[H]
\begin{center}
\end{center}
\centering
\begin{subfigure}{.4\textwidth}
  \centering
  \includegraphics[scale=0.40]{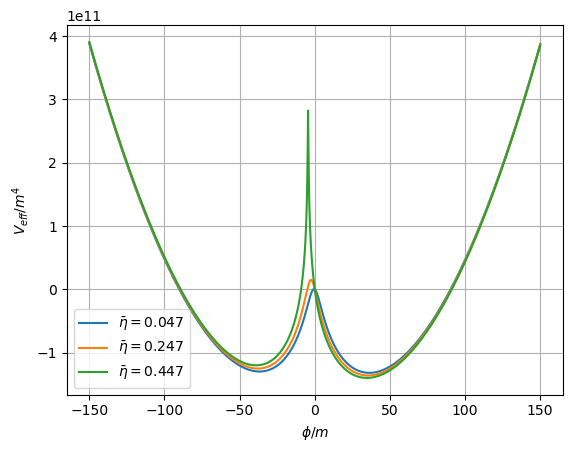}
  \caption{$\eta\phi^3 +\lambda\phi^4$ interaction, $\lambda=0.1$}
  \label{fig:sub7}
\end{subfigure}%
\begin{subfigure}{.7\textwidth}
  \centering
  \includegraphics[scale=0.40]{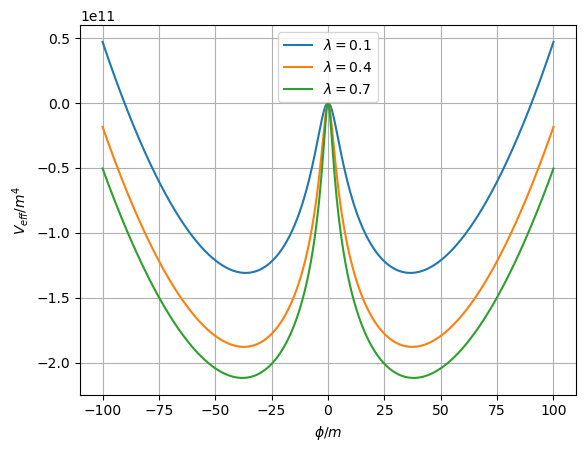}
  \caption{$\lambda\phi^4$ interaction}
  \label{fig:sub8}
\end{subfigure}
\caption{\it \small The one loop effective potential, \ref{v56}, at vanishing temperature in the de Sitter spacetime. Comparing the latter with \ref{fig:sub4}, \ref{fig:sub6}, we see that spontaneous symmetry breaking in  de Sitter with a massive scalar with $m^2{\ensuremath >}0$ is possible. In both the plots appearing above we have taken $\Lambda \sim 30 {\rm Gev}^{2}$, and $\xi \sim 0.13$. }
\label{fig:test4}
\end{figure}

\begin{figure}[H]
\begin{center}
\end{center}
\centering
\begin{subfigure}{.4\textwidth}
  \centering
  \includegraphics[scale=0.38]{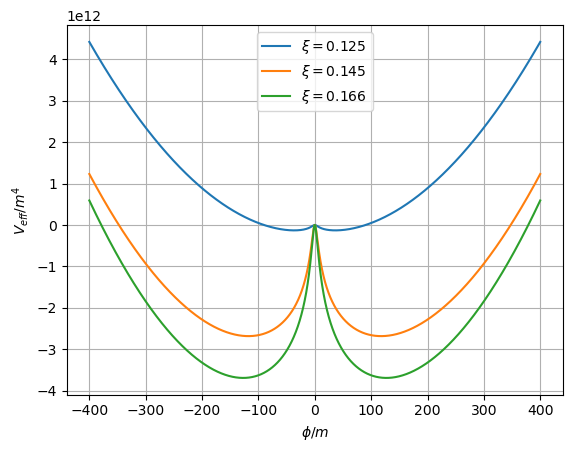}
  \caption{$\lambda\phi^4$ interaction}
  \label{fig:sub7}
\end{subfigure}%
\begin{subfigure}{.7\textwidth}
  \centering
  \includegraphics[scale=0.38]{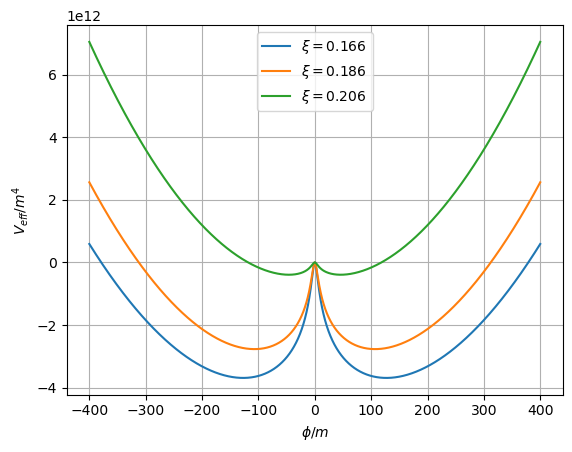}
  \caption{$\lambda\phi^4$ interaction}
  \label{fig:sub8}
\end{subfigure}
\caption{\it \small The one loop effective potential, \ref{v56}, at vanishing temperature in the de Sitter spacetime for different values of the non-minimal coupling  $\xi$. Note that for $\xi \sim 0.167$ or $1/6$, the dip or the hill of the effective potential is maximum. This effect originates from the terms appearing  on the last line on the right hand side of \ref{v56}. See main text for discussion. }
\label{fig:test4'}
\end{figure}

\begin{figure}[H]
\begin{center}
\end{center}
\centering
\begin{subfigure}{.4\textwidth}
  \centering
  \includegraphics[scale=0.38]{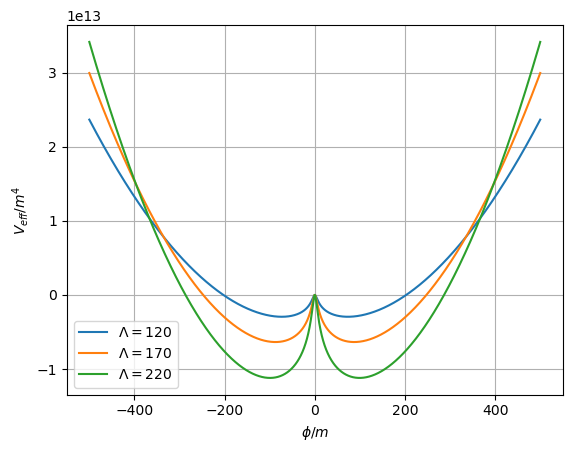}
  \caption{$\lambda\phi^4$ interaction}
  \label{fig:sub7}
\end{subfigure}%
\begin{subfigure}{.7\textwidth}
  \centering
  \includegraphics[scale=0.38]{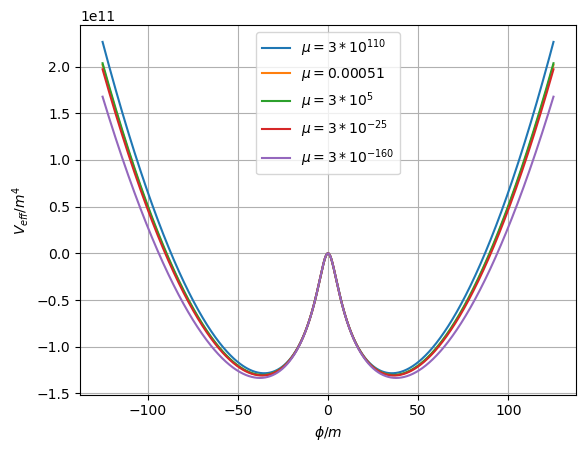}
  \caption{$\lambda\phi^4$ interaction}
  \label{fig:sub8}
\end{subfigure}
\caption{\it \small The one loop effective potential, \ref{v56}, with vanishing temperature in the de Sitter spacetime for different values of the cosmological constant  $\Lambda$, and the mass scale $\mu$ (in GeV). Increasing the value of $\Lambda$ increases the strength of the symmetry breaking effect. Note also that the symmetry breaking feature of the potential remains valid for a very wide range of the renormalisation  scale, $\mu$. }
\label{fig:test4''}
\end{figure}

\noindent
The metric for the cosmological de Sitter spacetime reads,
$$ds^2 =-dt^2 + e^{2\sqrt{\Lambda}t/\sqrt{3}} \left(dx^2 + dy^2 + dz^2 \right),$$
for which we have $R = 4\Lambda$, $R^{\mu\nu}R_{\mu\nu} =4 \Lambda^2$ and $R^{\mu\nu\rho\sigma}R_{\mu\nu\rho\sigma} = 8\Lambda^2/3$ in \ref{v30'},   $\Lambda$ being the positive cosmological constant. We have the one loop effective potential 
\begin{eqnarray}
&&\frac{V_{\rm{eff}}}{m^4} = \frac{\bar{\phi}^2 }{2}  + \frac{\bar{\eta}\bar{\phi}^3}{3!}  + \frac{\lambda\bar{\phi}^4}{4!} + \frac{2\xi \Lambda \bar{\phi}^2}{m^2} \nonumber\\&&  + 0.0015 \times \left[\left(2\left(\frac12\lambda \bar{\phi}^2 + \bar{\eta}\bar{\phi} \right)+\left(\frac12\lambda \bar{\phi}^2+\bar{\eta}\bar{\phi} \right)^2 \right) \left(\ln \frac{1+\frac12\lambda\bar{\phi}^2+\bar{\eta}\bar{\phi} }{\frac{4\pi \mu^2}{m^2}}-\frac12-\psi(2)\right) + \ln \left(1+\frac12\lambda \bar{\phi}^2+\bar{\eta}\bar{\phi}  \right)\right]  \nonumber\\&&  -0.012\times \frac{\left(\frac16 -\xi\right)\Lambda}{m^2} \left[\left(\frac12\lambda \bar{\phi}^2+\bar{\eta}\bar{\phi} \right) \left(\ln \frac{1+\frac12\lambda \bar{\phi}^2+\bar{\eta}\bar{\phi}}{\frac{4\pi \mu^2}{m^2}}  -\psi(2)\right)+\ln \left(1+\frac12\lambda \bar{\phi}^2+\bar{\eta}\bar{\phi} \right) \right] \nonumber\\&& +\frac{0.032 \Lambda^2}{m^4}\ln \left(1+\frac12\lambda \bar{\phi}^2+\bar{\eta}\bar{\phi}\right)\left[\left(\frac16 -\xi\right)^2-0.001\right].
\label{v56}
\end{eqnarray}
 The series appearing in the above equation might seem non-convergent for $\Lambda/m^2 \ll 1$. This  is simply an artefact of our scaling of $V_{\rm{eff}}$ by $m^4$. It is easy to see that this apparent ambiguity is absent, should  we instead make the scaling with respect to $\Lambda^2$.   However, since we are primarily  dealing with general curved spacetimes, we wish to keep this generic scaling with respect to $m^4$. We have investigated various features of this dimensionless effective potential in \ref{fig:test4}, \ref{fig:test4'} and \ref{fig:test4''}, for some customary numerical values of different parameters. In particular, we note that (with $\bar{\eta}=0$), spontaneous symmetry breaking at one loop is possible in de Sitter even with  $\xi, m^2{\ensuremath >} 0$. Usually, this has {\it no} analogue for flat or the Ricci flat spacetimes we have considered earlier, \ref{fig:sub4}, \ref{fig:sub6}. In fact symmetry breaking for  a massive scalar in flat spacetime seems only to be possible if the mass scale $\mu$ is extremely large compared to $m$. In de Sitter, this effect originates from the term appearing in the last line of the right hand side of \ref{v56}, which, when $\xi$ is $1/6$ or close to it, yields a negative contribution which can dominate the  $\bar{\phi}^2/2$ and other positive terms for small values of $\bar{\phi}$, leading to the symmetry breaking effective potential. It is easy to see that this effect will be present even with a tiny value of $\Lambda$. However, such curvature driven symmetry breaking may {\it a priori} be expected to be significant in the early inflationary universe scenario, where the dark energy density and hence the spacetime curvature was much higher compared to that of today. Note also from the second of \ref{fig:test4''} that symmetry breaking remains possible for a very wide range of the renormalisation scale, $\mu$.
 Before we proceed further, a couple of clarifying remarks  are in order here. First, we wish to emphasise that the Schwinger-DeWitt expansion formalism based upon the local Lorentz invariance  and four momentum we are using here, is essentially meant to capture the curvature effects to ultraviolet quantum processes, such as the anomaly~\cite{Toms}. Being ultraviolet, these are  short scale or local effects. Thus, the current formalism, being based upon the local geometry,  is not suitable to address the late time, deep infrared or super-Hubble phenomenon like the non-perturbative secular effect in de Sitter~\cite{Prokopec:2002uw, Krotov:2010ma, Akhmedov:2013vka, Wang:2022mvv, Sadekov:2023ivd, Akhmedov:2024npw, Bhattacharya:2023yhx} (also references therein) for a light scalar. One needs to use the exact propagator and the Schwinger-Keldysh closed time path formalism to address such non-equilibrium scenarios. Thus the effect we are looking into could be some initial or transient phenomenon, with time scale much small compared to that of the duration of the inflation.   Thus the present formalism can be thought of as somewhat analogous to the standard particle physics set up, where we essentially ignore the global structure of the spacetime.  In fact standard particle physics results or techniques including in-out scattering without considering {\it any} curvature have been applied to early inflationary cosmology earlier, provided they are very high energy and short ranged in spacetime, e.g.~\cite{Linde:1990flp, Mazumdar:2018dfl} and references therein. Nevertheless, after the symmetry breaking occurs at short scale, it will be interesting to look into its effect on the background field's  classical or stochastic behaviour at late times and at large scale, from the effective action corresponding to \ref{v56}.

 Finally, we  note that had we ignored the quadratic curvature terms from the beginning, there would be no such SSB described above. Will the result still be valid, should we include cubic order curvature terms? Putting things together now, we believe this feature, along with the thermal self energy result \ref{g4}, sufficiently motivates us to look into this problem via some suitable non-perturbative resummation techniques. Any such technique is not only required to  resum over the coupling, but also to resum some curvature terms as well, to see its effect on the short scale physics we are probing here.  We reserve these problems for a future work.   However in this paper, as a bridge between the above one loop result and a non-perturbative analysis, we wish to present some two loop computations for the effective potential. Our motivation is to see whether the above result gets any qualitative modification at higher loop order. Indeed as we will see, there is no qualitative deviation from SSB in de Sitter. Another motivation of us is to  check, in the context of finite temperature field theory, that the field renormalisation counterterms are indeed independent of the spacetime curvature.   

We now wish to discuss the above effective potentials in the context of finite temperature field theory.

%%%%%%%%%%%%%%%%%%%%%%%%%%%
\subsection{Phase transition at finite temperature}\label{S8}\
%%%%%%%%%

\noindent
\underline{\bf $m^2{\ensuremath <}0$ case :}\\\\

\begin{figure}[H]
\begin{center}
\includegraphics[scale=0.40]{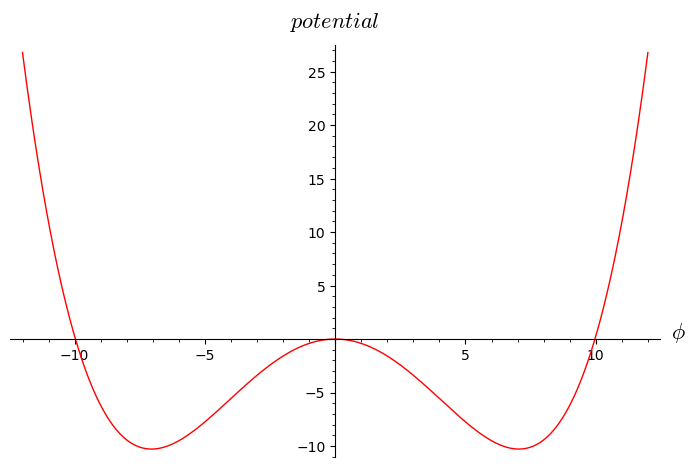}
\caption{  \it \small Tree level symmetry breaking potential with  $m^2<0$ and positive quartic self interaction.}
\label{symb1}
\end{center}
\end{figure}

\begin{figure}[H]
\begin{center}
\end{center}
\centering
\begin{subfigure}{.4\textwidth}
  \centering
  \includegraphics[scale=0.4]{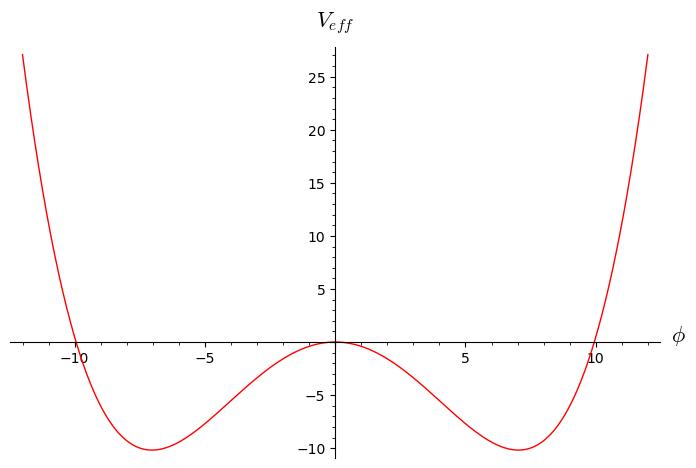}
  \caption{\it \small Symmetry breaking with the effective potential of \ref{v58}. Here $T_c \sim 14.77{\rm GeV}$ and $T \sim 1{\rm GeV}$. }
  \label{fig:sub9}
\end{subfigure}%
\begin{subfigure}{.7\textwidth}
  \centering
  \includegraphics[scale=0.4]{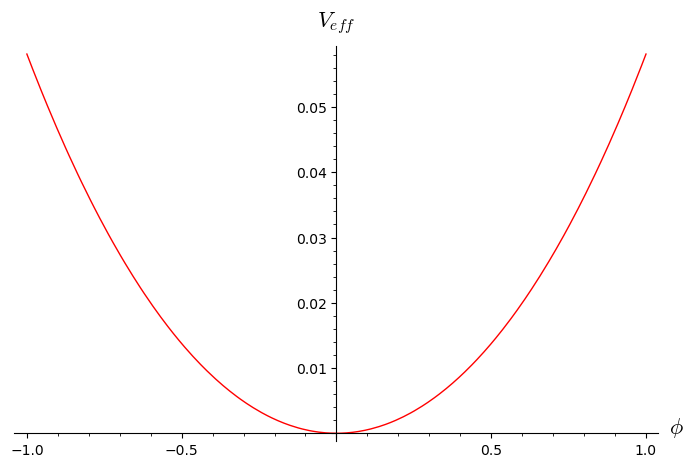}
  \caption{\small \it Symmetry restoration with $T>T_{c}$.}
  \label{fig:sub10}
\end{subfigure}
\caption{\it \small Symmetry breaking and restoration of the thermal effective potential of \ref{v58} in flat spacetime with $m^2{\ensuremath <} 0$. }
\label{fig:test5}
\end{figure}

\noindent
To begin with, let us consider the thermal effective potential \ref{v46} at zero spacetime curvature with only quartic coupling and at high temperature, e.g.~\cite{Nastase:1970yyp}, 
\begin{eqnarray}
V_{\rm{eff},\beta} =  \frac{m^{2} \phi^2}{2}   + \frac{\lambda\phi^4}{4!}  + \left[-\frac{ \pi^2}{90\beta^4}+\frac{m^{2}+\frac12 \lambda\phi^2}{24\beta^2}\right] \qquad (\lambda >0),
\label{v57}
\end{eqnarray}
where we have kept only the first two thermal terms for simplicity.  It is well known that for $m^2{\ensuremath <}0$ there can be  symmetry breaking for the tree level potential, as has been depicted in \ref{symb1}.
Following e.g.~\cite{Nastase:1970yyp}, we now rewrite \ref{v57} as
\begin{eqnarray}
V_{\rm{eff},\beta} =  \frac{m^{2} \phi^2}{2}  \left(1 - \frac{T^2}{T^{2}_{c}}\right)  + \frac{ \lambda\phi^4 }{4!},
\label{v58}
\end{eqnarray}
where, $T_c = \sqrt{-\frac{24m^2}{\lambda}}$ is called the critical temperature, and we have ignored the two $\phi$-independent terms.  Since $m^2{\ensuremath<}0$,  there is always a  symmetry breaking for $T {\ensuremath <} T_c$, but for $T \geq T_c$, there can be no such thing. This indicates that  a  second-order phase transition can occur here. We have depicted the scenario in \ref{fig:test5}.
Let us now add the tree level curvature term $\xi R \phi^2/2$ to see if there is any modification to the above phenomenon. We have 
\begin{eqnarray}
&&V_{\rm{eff},\beta}= \frac{m^{2} \phi^2}{2}  \left(1 - \frac{T^2}{T^{2}_{c}} +\frac{\xi R}{m^2}\right)+ \frac{ \lambda\phi^4 }{4!}.
\label{v59}
\end{eqnarray}
This indicates a new critical temperature, say $T_{1c}$, given by
\begin{eqnarray}
 T_{1c}=\left(1+\frac{\xi R}{m^2}\right)^{\frac12}T_{c}, 
\label{v60}
\end{eqnarray}
showing that for $\xi R {\ensuremath >}0$, a reduction in the value of the  critical temperature and an increment for the opposite sign.

\begin{figure}[H]
\begin{center}
\includegraphics[scale=0.50]{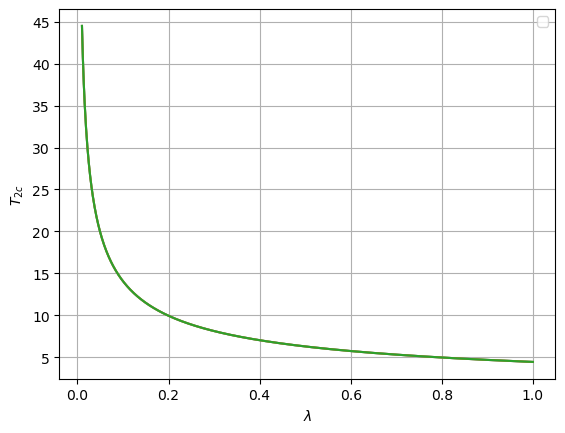}
\caption{\it \small Variation of the effective critical temperature with respect to $\lambda$, \ref{v62} in the de Sitter spacetime.  We have taken $\xi \sim 0.125$ and $\Lambda \sim 30{\rm GeV}^{2}$.}
\label{symmb2}
\end{center}
\end{figure}

At one loop, for $\beta \to 0 $, we get after keeping only terms linear in the spacetime curvature,
\begin{eqnarray}
V_{\rm{eff,\beta}}=\frac{\lambda\phi^4}{4!}+\frac{m^2\phi^2}{2}\left[1-\frac{T^2}{T^{2}_{c}}+\frac{\xi R}{m^2} -\frac{3 \lambda}{64 \pi^2} - 0.732 \times \frac{\left(\frac16 -\xi\right)R \lambda}{16\pi^2 m^2}\right].
\label{v61}
\end{eqnarray}
Now the critical temperature reads
\begin{eqnarray}
&&T_{1c}=\left[1+\frac{\xi R}{m^2}-\frac{3 \lambda}{64 \pi^2} - 0.732 \times \frac{\left(\frac16 -\xi\right)R \lambda}{16\pi^2 m^2}\right]^{\frac12}T_c.
\label{v62}
\end{eqnarray}
For the de Sitter spacetime, $(R=4\Lambda)$, we have plotted the variation of $T_{1c}$ with respect to the quartic coupling  in \ref{symmb2}.
This completes our brief discussion on the case of negative rest mass squired of the scalar field. We wish to discuss below the usual case when $m^2$ is positive.\\

\noindent
\underline{\bf $m^2>0$ case :}\\
%%%%%

\noindent
Since \ref{v46} is essentially a high temperature expansion, and we are not setting the mass scale to be equal to $\beta^{-1}$, we will ignore the zero temperature part in \ref{v46}. In particular, we have explicitly verified that even if we keep them, they bring in no qualitative changes to the symmetry restoration phenomenon at high temperature. Also, in the finite temperature part of the effective action,  we have not included the odd power terms of $m_1$, as they originate from the zero mode fluctuations (i.e. the $n=0$ term in \ref{v36}). We refer our reader to~\cite{dolan, Hu} and references therein in this context. Nevertheless, we also have explicitly verified that for $\bar{\eta}=0$, inclusion of such terms does not alter any qualitative feature of the symmetry restoration phenomenon. 

\begin{figure}[H]
\begin{center}
\end{center}
\centering
\begin{subfigure}{.4\textwidth}
  \centering
  \includegraphics[scale=0.3]{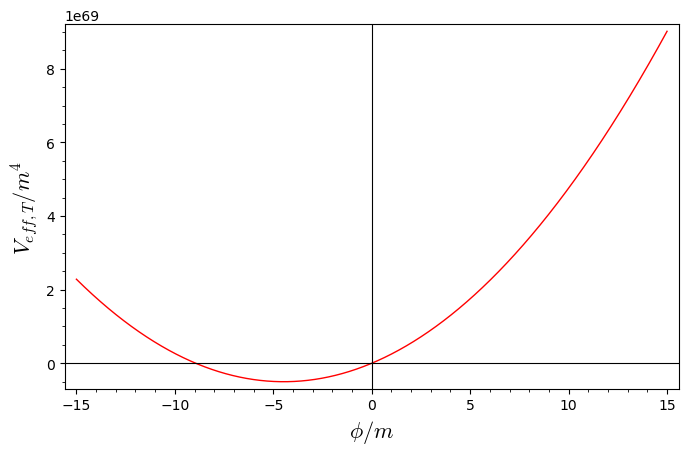}
  \caption{One loop $\eta\phi^3+\lambda\phi^4$ interaction }
  \label{fig:sub11}
\end{subfigure}%
\begin{subfigure}{.7\textwidth}
  \centering
  \includegraphics[scale=0.3]{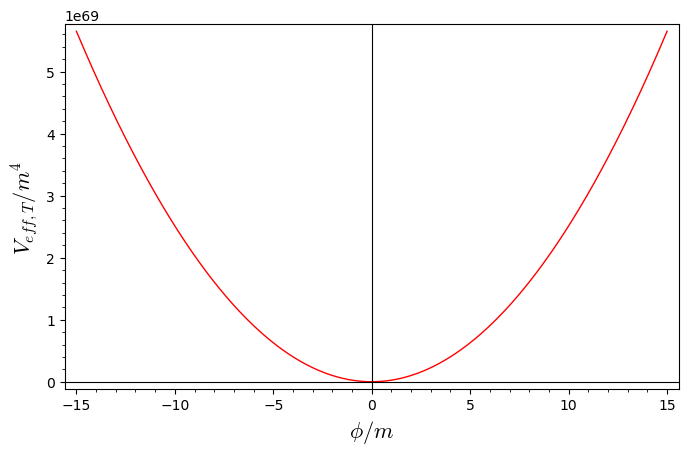}
  \caption{One loop $\lambda\phi^4$ interaction}
  \label{fig:sub12}
\end{subfigure}
\label{fig-flatth}
\caption{\it \small Plot of the effective  potential at finite temperature $(\beta=10^{-5} \rm{GeV^{-1}},\eta=0.447,\lambda=0.1)$ with vanishing background curvature, \ref{v64}. We have not considered finite terms independent of $\phi$. See main text for discussion.} 
\label{fig:test6}
\end{figure}

Let us accordingly begin with the thermal effective potential with zero background spacetime curvature found from \ref{v46},
\begin{eqnarray}
&&\frac{V_{\rm{eff},\beta}}{m^4} = \frac{\bar{\phi}^2}{2}  + \frac{\bar{\eta}\bar{\phi}^3}{3!}  + \frac{\lambda\bar{\phi}^4}{4!} + \left[\frac{\frac12\lambda\bar{\phi}^2+\bar{\eta}\bar{\phi}}{24m^2\beta^2}      -0.001\times\left(2\left(\frac12\lambda\bar{\phi}^2+\bar{\eta}\bar{\phi}\right)+\left(\frac12\lambda\bar{\phi}^2+\bar{\eta}\bar{\phi}\right)^2\right) \ln {\left((m\beta)^2 \left(1+\frac12\lambda\bar{\phi}^2+\bar{\eta}\bar{\phi}\right)\right)}\right. \nonumber\\&& \left. +0.00004\times (m\beta)^2\left(3\left(\frac12\lambda\bar{\phi}^2+\bar{\eta}\bar{\phi}\right)+3\left(\frac12\lambda\bar{\phi}^2+\bar{\eta}\bar{\phi}\right)^2+\left(\frac12\lambda\bar{\phi}^2+\bar{\eta}\bar{\phi}\right)^3\right) \right. \nonumber\\&& \left. -4.37\times 10^{-8}(m\beta)^4 \left(6\left(\frac12\lambda\bar{\phi}^2+\bar{\eta}\bar{\phi}\right)^2+4\left(\frac12\lambda\bar{\phi}^2+\bar{\eta}\bar{\phi}\right)+4\left(\frac12\lambda\bar{\phi}^2+\bar{\eta}\bar{\phi}\right)^3+\left(\frac12\lambda\bar{\phi}^2+\bar{\eta}\bar{\phi}\right)^4\right) \right. \nonumber\\&& \left. +5.38\times 10^{-10}(m\beta)^6\left(5\left(\frac12\lambda\bar{\phi}^2+\bar{\eta}\bar{\phi}\right)+10\left(\frac12\lambda\bar{\phi}^2+\bar{\eta}\bar{\phi}\right)^2+10\left(\frac12\lambda\bar{\phi}^2+\bar{\eta}\bar{\phi}\right)^3+5\left(\frac12\lambda\bar{\phi}^2+\bar{\eta}\bar{\phi}\right)^4+\left(\frac12\lambda\bar{\phi}^2+\bar{\eta}\bar{\phi}\right)^5      \right)\right],
\nonumber\\
\label{v64}
\end{eqnarray}
which we have plotted in \ref{fig:test6},  ignoring the $\phi$-independent finite terms. Comparing this with \ref{fig:test2} for zero temperature, we see no qualitative changes. This is expected, as if there is no symmetry breaking at zero temperature, one usually does not expect the same at finite temperature. 

\begin{figure}[H]
\begin{center}
\end{center}
\centering
\begin{subfigure}{.4\textwidth}
  \centering
  \includegraphics[scale=0.24]{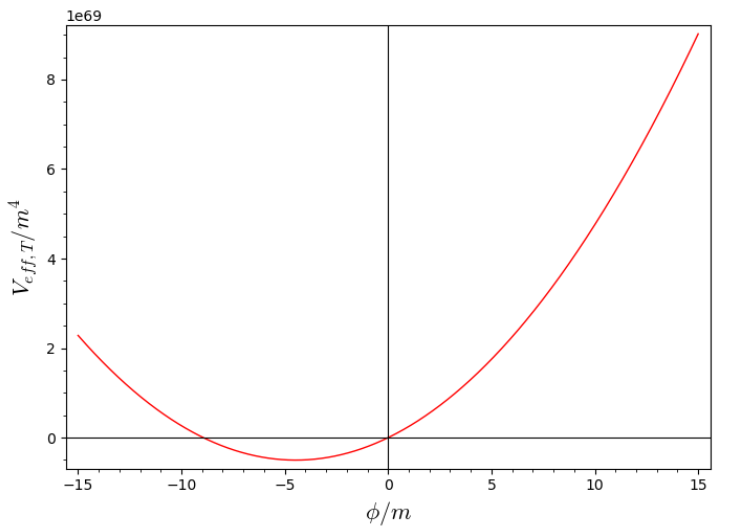}
  \caption{One loop $\eta\phi^3+\lambda\phi^4$ interaction (Schwarzschild) }
  \label{fig:sub13}
\end{subfigure}%
\begin{subfigure}{.7\textwidth}
  \centering
  \includegraphics[scale=0.24]{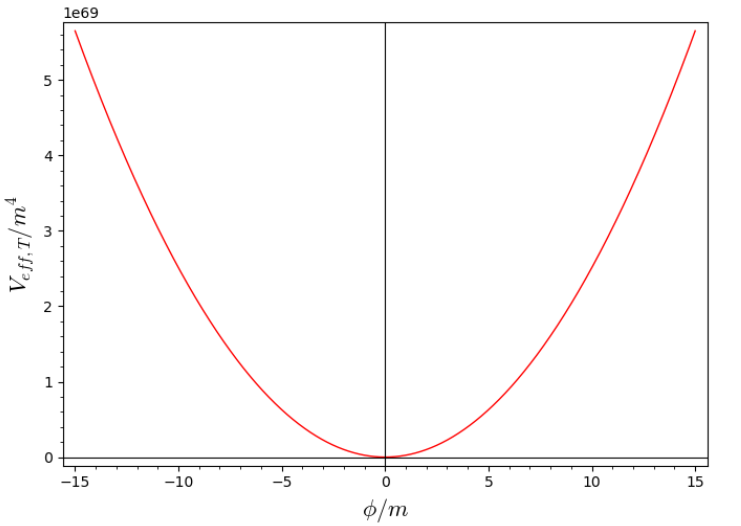}
  \caption{One loop $\lambda\phi^4$ interaction (Schwarzschild)}
  \label{fig:sub14}
\end{subfigure}
\caption{\it \small Variation of the effective potential for one loop at finite temperature for the Schwarzschild spacetime, \ref{v65}.}
\label{fig:test7}
\end{figure}

\noindent
For the Schwarzschild spacetime, we have  from \ref{v46}
\begin{eqnarray}
&&\frac{V_{\rm{eff},\beta}}{m^4} = \frac{\bar{\phi}^2}{2}  + \frac{\bar{\eta}\bar{\phi}^3}{3!}  + \frac{\lambda\bar{\phi}^4}{4!} + \left[\frac{1}{24}\frac{\frac12\lambda\bar{\phi}^2+\bar{\eta}\bar{\phi}}{m^2\beta^2}     -0.001\times \left(2\left(\frac12\lambda\bar{\phi}^2+\bar{\eta}\bar{\phi}\right)+\left(\frac12\lambda\bar{\phi}^2+\bar{\eta}\bar{\phi}\right)^2\right) \ln \left((m\beta)^2 \left(1+\frac12\lambda\bar{\phi}^2+\bar{\eta}\bar{\phi}\right)\right) \right. \nonumber\\&& \left. +0.00004\times (m\beta)^2\left(3\left(\frac12\lambda\bar{\phi}^2+\bar{\eta}\bar{\phi}\right)+3\left(\frac12\lambda\bar{\phi}^2+\bar{\eta}\bar{\phi}\right)^2+\left(\frac12\lambda\bar{\phi}^2+\bar{\eta}\bar{\phi}\right)^3\right) \right. \nonumber\\&& \left. -4.37\times 10^{-8}(m\beta)^4 \left(6\left(\frac12\lambda\bar{\phi}^2+\bar{\eta}\bar{\phi}\right)^2+4\left(\frac12\lambda\bar{\phi}^2+\bar{\eta}\bar{\phi}\right)+4\left(\frac12\lambda\bar{\phi}^2+\bar{\eta}\bar{\phi}\right)^3+\left(\frac12\lambda\bar{\phi}^2+\bar{\eta}\bar{\phi}\right)^4\right) \right. \nonumber\\&& \left. +5.38\times 10^{-10}(m\beta)^6\left(5\left(\frac12\lambda\bar{\phi}^2+\bar{\eta}\bar{\phi}\right)+10\left(\frac12\lambda\bar{\phi}^2+\bar{\eta}\bar{\phi}\right)^2+10\left(\frac12\lambda\bar{\phi}^2+\bar{\eta}\bar{\phi}\right)^3+5\left(\frac12\lambda\bar{\phi}^2+\bar{\eta}\bar{\phi}\right)^4+\left(\frac12\lambda\bar{\phi}^2+\bar{\eta}\bar{\phi}\right)^5      \right)\right]  \nonumber\\&&- \frac{0.0002}{m^4 G^4 M^4} \left[ -0.003\times (m\beta)^2\left(\frac12\lambda\bar{\phi}^2+\bar{\eta}\bar{\phi}\right) +0.00004 \times (m\beta)^4\left(2\left(\frac12\lambda\bar{\phi}^2+\bar{\eta}\bar{\phi}\right)+\left(\frac12\lambda\bar{\phi}^2+\bar{\eta}\bar{\phi}\right)^2\right)\right. \nonumber\\&& \left. +0.00001\times (m\beta)^6\left(3\left(\frac12\lambda\bar{\phi}^2+\bar{\eta}\bar{\phi}\right)+3\left(\frac12\lambda\bar{\phi}^2+\bar{\eta}\bar{\phi}\right)^2+\left(\frac12\lambda\bar{\phi}^2+\bar{\eta}\bar{\phi}\right)^3\right)\right],
\label{v65}
\end{eqnarray}
which we have plotted in \ref{fig:test7} for $GM \sim 10^{2}{\rm m}$. Comparing them with the results of \ref{S6}, we see, expectedly, no qualitatively new feature is added. \\

\begin{figure}[H]
\begin{center}
\end{center}
\centering
\begin{subfigure}{.4\textwidth}
  \centering
  \includegraphics[scale=0.24]{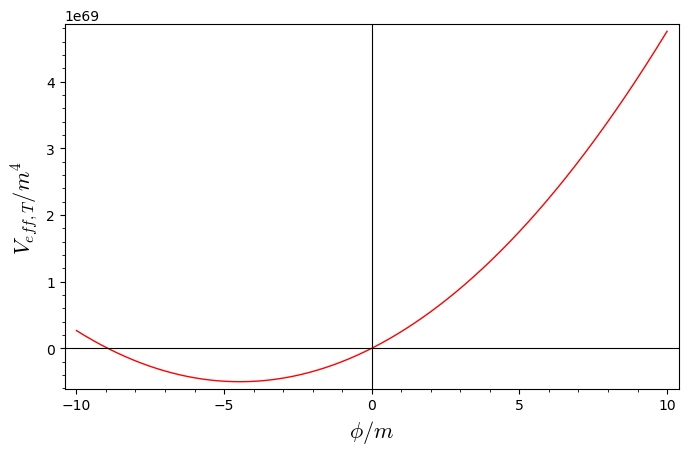}
  \caption{One loop $\eta\phi^3+\lambda\phi^4$ interaction (de Sitter)}
  \label{fig:sub15}
\end{subfigure}%
\begin{subfigure}{.7\textwidth}
  \centering
  \includegraphics[scale=0.24]{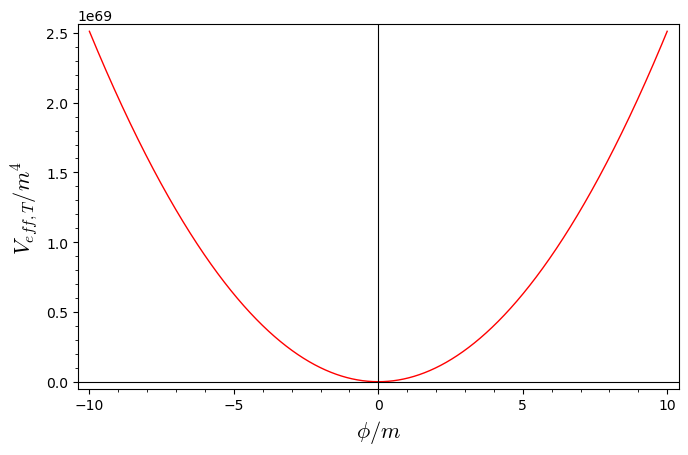}
  \caption{One loop $\lambda\phi^4$ interaction (de Sitter)}
  \label{fig:sub16}
\end{subfigure}
\caption{\it \small Plot of the finite temperature effective  potential in the de Sitter background ($\beta=10^{-5} \rm{GeV^{-1}},\eta=0.447,\lambda=0.1,\Lambda \sim 30{\rm GeV}^{2}$ and $\xi = 0.125$). Comparing this with the zero temperature result of \ref{S6}, we see symmetry restoration effect at finite temperature and hence indication of a second order phase transition.}
\label{fig:test8}
\end{figure}

\noindent
Finally, for the de Sitter spacetime, we have
{\small
\begin{eqnarray}
&&\frac{V_{\rm{eff},\beta}}{m^4} = \frac{\bar{\phi}^2}{2}   + \frac{\bar{\eta}\bar{\phi}^3}{3!}  + \frac{\lambda\bar{\phi}^4}{4!}+\frac{2\xi\Lambda \bar{\phi}^2}{m^2}+ \left[\frac{\frac12\lambda\bar{\phi}^2+\bar{\eta}\bar{\phi}}{24m^2\beta^2}     -0.001 \times \left(2\left(\frac12\lambda\bar{\phi}^2+\bar{\eta}\bar{\phi}\right)+\left(\frac12\lambda\bar{\phi}^2+\bar{\eta}\bar{\phi}\right)^2\right) \ln \left((m\beta)^2 \left(1+\frac12\lambda\bar{\phi}^2+\bar{\eta}\bar{\phi}\right)\right)  \right. \nonumber\\&& \left. +0.00004\times (m\beta)^2\left(3\left(\frac12\lambda\bar{\phi}^2+\bar{\eta}\bar{\phi}\right)+3\left(\frac12\lambda\bar{\phi}^2+\bar{\eta}\bar{\phi}\right)^2+\left(\frac12\lambda\bar{\phi}^2+\bar{\eta}\bar{\phi}\right)^3\right)\right. \nonumber\\&& \left. -4.37\times 10^{-8}(m\beta)^4 \left(6\left(\frac12\lambda\bar{\phi}^2+\bar{\eta}\bar{\phi}\right)^2+4\left(\frac12\lambda\bar{\phi}^2+\bar{\eta}\bar{\phi}\right)+4\left(\frac12\lambda\bar{\phi}^2+\bar{\eta}\bar{\phi}\right)^3+\left(\frac12\lambda\bar{\phi}^2+\bar{\eta}\bar{\phi}\right)^4\right) \right. \nonumber\\&& \left. +5.38\times 10^{-10}(m\beta)^6\left(5\left(\frac12\lambda\bar{\phi}^2+\bar{\eta}\bar{\phi}\right)+10\left(\frac12\lambda\bar{\phi}^2+\bar{\eta}\bar{\phi}\right)^2+10\left(\frac12\lambda\bar{\phi}^2+\bar{\eta}\bar{\phi}\right)^3+5\left(\frac12\lambda\bar{\phi}^2+\bar{\eta}\bar{\phi}\right)^4+\left(\frac12\lambda\bar{\phi}^2+\bar{\eta}\bar{\phi}\right)^5      \right) \right]\nonumber\\&&-0.12\left(\frac16 -\xi\right)\frac{\Lambda}{m^2} \left[-0.1 \ln \left(m^2\beta^2 \left(1+\frac12\lambda\bar{\phi}^2+\bar{\eta}\bar{\phi}\right)\right)  +0.001\times (m\beta)^2 \left(2\left(\frac12\lambda\bar{\phi}^2+\bar{\eta}\bar{\phi}\right)+\left(\frac12\lambda\bar{\phi}^2+\bar{\eta}\bar{\phi}\right)^2\right) \right. \nonumber\\&& \left. +0.00001 \times (m\beta)^4\left(3\left(\frac12\lambda\bar{\phi}^2+\bar{\eta}\bar{\phi}\right)+3\left(\frac12\lambda\bar{\phi}^2+\bar{\eta}\bar{\phi}\right)^2+\left(\frac12\lambda\bar{\phi}^2+\bar{\eta}\bar{\phi}\right)^3\right)\right. \nonumber\\&& \left. + 0.00002\times (m\beta)^6\left(6\left(\frac12\lambda\bar{\phi}^2+\bar{\eta}\bar{\phi}\right)^2+4\left(\frac12\lambda\bar{\phi}^2+\bar{\eta}\bar{\phi}\right)+4\left(\frac12\lambda\bar{\phi}^2+\bar{\eta}\bar{\phi}\right)^3+\left(\frac12\lambda\bar{\phi}^2+\bar{\eta}\bar{\phi}\right)^4\right)\right]  \nonumber\\&&- \frac{0.03}{m^4}\left(\left(\frac16 -\xi\right)^2 16 \Lambda^2 - \frac{2}{135} \Lambda^2 \right)\left[ -0.003 \times (m\beta)^2\left(\frac12\lambda\bar{\phi}^2+\bar{\eta}\bar{\phi}\right) +0.00004\times (m\beta)^4\left(2\left(\frac12\lambda\bar{\phi}^2+\bar{\eta}\bar{\phi}\right)+\left(\frac12\lambda\bar{\phi}^2+\bar{\eta}\bar{\phi}\right)^2\right)\right. \nonumber\\&& \left. +0.00001 \times (m\beta)^6\left(3\left(\frac12\lambda\bar{\phi}^2+\bar{\eta}\bar{\phi}\right)+3\left(\frac12\lambda\bar{\phi}^2+\bar{\eta}\bar{\phi}\right)^2+\left(\frac12\lambda\bar{\phi}^2+\bar{\eta}\bar{\phi}\right)^3\right)\right]+ 0.24 \times \left(\frac16 -\xi\right)^2   \frac{\Lambda^2}{m^4}\left[-0.003 \times (m\beta)^2\left(\frac12\lambda\bar{\phi}^2+\bar{\eta}\bar{\phi}\right)\right. \nonumber\\&& \left. +0.00004\times (m\beta)^4\left(2\left(\frac12\lambda\bar{\phi}^2+\bar{\eta}\bar{\phi}\right)+\left(\frac12\lambda\bar{\phi}^2+\bar{\eta}\bar{\phi}\right)^2\right) +0.00001\times (m\beta)^6\left(3\left(\frac12\lambda\bar{\phi}^2+\bar{\eta}\bar{\phi}\right)+3\left(\frac12\lambda\bar{\phi}^2+\bar{\eta}\bar{\phi}\right)^2+\left(\frac12\lambda\bar{\phi}^2+\bar{\eta}\bar{\phi}\right)^3\right)\right].
\nonumber\\
\label{v66}
\end{eqnarray}}
The variation of the above effective potential can be seen in \ref{fig:test8}, after ignoring $\phi$-independent finite terms. We have taken $\Lambda \sim 30{\rm GeV}^{2}$ and $\xi = 0.125$. Comparing this with the zero temperature result of \ref{fig:test4}, we see symmetry restoration. This must be  a second order phase transition.
 The above results are essentially valid at high temperatures. With a low temperature effective potential, we may actually see how the broken symmetry is restored at some critical temperature. Following~\cite{dolan, Nastase:1970yyp}, we keep only the ${\cal O}(\beta^{-2})$ term in this case, so that for de Sitter, we obtain a low temperature version of $V_{{\rm eff},\beta}$,
\begin{figure}[H]
\begin{center}
\includegraphics[scale=0.50]{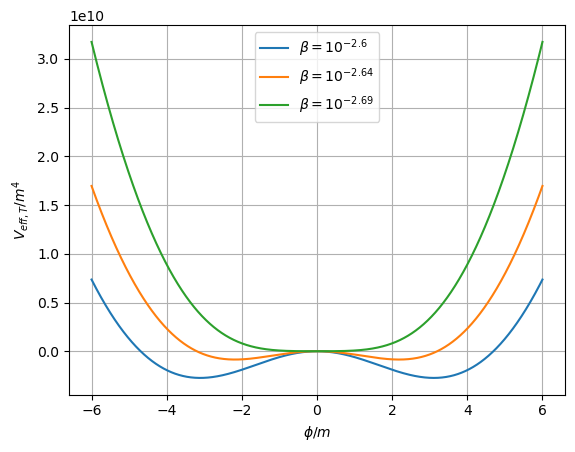}
\caption{\it \small  Variation of the low temperature effective potential, \ref{2R1}, with respect to $\bar{\phi}=\phi/m$, for various values of $\beta$.  We have taken $\xi \sim 0.125$ and $\Lambda \sim 30{\rm GeV}^{2}$. Symmetry restoration and the existence of a critical temperature can be explicitly seen here.}
\label{symmb3}
\end{center}
\end{figure}
\be
V_{{\rm eff},\beta}\simeq \frac12 m^2 \phi^2+2\xi\Lambda \phi^2+0.032\Lambda^2 \ln\left(1+\frac{\lambda {\phi}^2}{2m^2}\right)\left[\left(\frac16 -\xi\right)^2-0.001\right]+\frac{\lambda \phi^2}{48\beta^2}.
\label{2R1}
\ee
For the sake of calculational simplicity, we further take $\phi/m \ll 1$. Then similar steps which led to \ref{v62} now yields the critical temperature,

% $$\frac{\bar{\phi^2}}{2}\left(1+4\xi\Lambda+\frac{0.032\lambda\Lambda^2}{m^4}\left(\left(\frac16 -\xi\right)^2-0.001\right)+\frac{\lambda}{24m^2\beta^2}\right).$$From here we can show that the critical temperature is 
%
\begin{eqnarray}
T_c=4.9 \sqrt{\frac{\Lambda}{\lambda}}\left[-\frac{m^2}{\Lambda}-4\xi+\frac{0.032\lambda \Lambda}{m^2}\left(0.001-\left(\frac16 -\xi\right)^2\right)\right]^{1/2},
\label{v66'}
\end{eqnarray} 
depicted in \ref{symmb3}, which is qualitatively similar to that of the $m^2\ensuremath{<}0$ case, discussed in \ref{S8}. It is also clear that a more general analysis requires writing the thermal propagators at general temperatures.

%%%%%%%%
\section{Two loop effective potential in curved spacetime}\label{S9}
%%%%%%%%

\begin{figure}[H]
\begin{center}
\end{center}
\centering
\begin{subfigure}{.25\textwidth}
  \centering
  \includegraphics[scale=0.5]{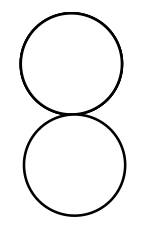}
  \caption{Double bubble}
  \label{fig:sub17}
\end{subfigure}%
\begin{subfigure}{.25\textwidth}
  \centering
  \includegraphics[scale=0.5]{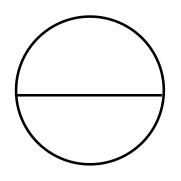}
  \caption{Sunset}
  \label{fig:sub18}
\end{subfigure}
\begin{subfigure}{.25\textwidth}
  \centering
  \includegraphics[scale=0.3]{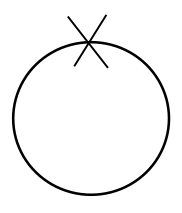}
  \caption{One loop counterterm}
  \label{fig:sub9}
\end{subfigure}
\caption{\it \small Two loop 1PI vacuum diagrams for $\lambda \phi^4+\eta \phi^3$ interactions. }
\label{fig:test}
\end{figure}

\noindent
We now wish to extend some of our earlier results to two loop order.  As we have stated towards the end of \ref{S6}, the following calculations are supposed to be a bridge between the one loop results and the non-perturbative computations we wish to do in future. Also, we wish to verify and demonstrate that the field renormalisation counterterms are indeed independent of the spactime curvature.  The two loop renormalisation of $\phi^4$-theory in general curved spacetimes was first done in \cite{bunch}. We note that at two loop, there can be non-local contributions from the integrals of the sunset diagram, \ref{fig:sub18}. This originates from the integration  of the non-local part of the self energy over the entire spacetime. However, as we have discussed, since the current formalism we are using is essentially based upon local Lorentz invariance and hence suitable at small scales only, any such non-local results could be misleading. Therefore we wish to focus only on the local contributions from the vacuum graphs. 

The effective action at this order reads ($\hbar =1$)
\be
- \frac{\lambda}{8} \int  \sqrt{-g}d^d x   G^2 (x,x)+\frac{1}{12}\int \sqrt{-g} \sqrt{-g'} d^d x d^d x' (\eta +\lambda \phi)^2   G^3 (x,x')+\ {\rm counterterm ~ contributions}.
\label{EA}
\ee
Note that the second or the sunset integral is quadratic in the couplings. Thus we may expect {\it a priori} that the first, i.e. the double bubble integral to make the leading contributions at two loop, also supported by our earlier argument of \ref{add}.  Note also that the double bubble yields a purely local contribution.
The corresponding 1PI vacuum diagrams can be seen \ref{fig:test}.
At zero temperature, computation of the divergent part of the effective action and subsequent renormalisation can be seen in~\cite{Toms}. We also refer our reader to~\cite{Odin}  for a renormalisation group derivation of a two loop effective potential with quartic self interaction under weak curvature approximation at zero temperature.\\

\noindent
Let us now compute the diagrams of \ref{fig:test}.
 In order to keep our expressions simple, we shall also make a linear in curvature approximation in the following. We  evaluate the zero temperature effective potential first, show the renormalisation  and then turn on the  temperature. We will see that the latter is divergence free, as expected. Finally, we shall present the finite part of the ${\cal O}(\lambda)$ double bubble diagram at finite temperature up to quadratic order in the curvature.    \\

\noindent
{\underline{\bf The double bubble}}:\\

\noindent
The double bubble diagram is given by the first of \ref{fig:test}. We have at 
 $d=4-\epsilon$,
\begin{eqnarray}
&&-\frac{\lambda}{8}\left(\int \frac{d^dk}{(2\pi)^d}\left(\frac{1}{k^2+m^{2}_{1}}+\frac{\left(\frac16 -\xi\right)R}{(k^2+m^{2}_{1})^2}\right)\right)^2= -\frac{\lambda m^{4}_{1}}{2(4\pi)^4}\left(\frac{\mu^{-2\epsilon}}{\epsilon^2}-\frac{\mu^{-\epsilon}}{\epsilon}\left(\ln{\frac{m^{2}_{1}}{4\pi\mu^2}}-\psi(2)\right)-\psi(2)\ln{\frac{m^{2}_{1}}{4\pi\mu^2}}\right. \nonumber\\&& \left. +\frac12\left(\ln{\frac{m^{2}_{1}}{4\pi\mu^2}}\right)^2 +\frac14\left(\frac{\pi^2}{3}+2\psi^2(2)-\psi^{\prime}(2)\right)\right) + \frac{\lambda\left(\frac16 -\xi\right)Rm^{2}_{1}}{4(4\pi)^4}\left(\frac{4 \mu^{-2\epsilon}}{\epsilon^2} +\frac{2\mu^{-\epsilon}}{\epsilon}\left(\psi(1)+\psi(2)\right)-\frac{4\mu^{-\epsilon}}{\epsilon}\ln{\frac{m^{2}_{1}}{4\pi\mu^2}}\right. \nonumber\\&& \left. +2\left(\ln{\frac{m^{2}_{1}}{4\pi\mu^2}}\right)^2 +\psi(1)\psi(2)+\frac12\left(\frac{\pi^2}{3}+\psi^2(2)-\psi^{\prime}(2)\right) +\frac12\left(\frac{\pi^2}{3}+\psi^2(1)-\psi^{\prime}(1)\right)+2\ln{\frac{4\pi\mu^2}{m^{2}_{1}}}\left(\psi(1)+\psi(2)\right)\right), \nonumber\\
\label{w1}
\end{eqnarray}
where as we have stated earlier, we have kept terms only linear in curvature and 
$$\psi^{\prime}(x)= \frac{\Gamma^{\prime\prime}(x)}{\Gamma(x)}-\frac{(\Gamma^{\prime}(x))^2}{\Gamma^2(x)}.$$

\noindent
{\underline{\bf The sunset}}:\\

\noindent
 As we have mentioned at the beginning of this section, for the sunset diagram we will consider only its local part, as any non-local part could yield misleading results in this local or UV effective formalism. The corresponding integral, up to linear in curvature terms reads
\begin{eqnarray}
&&\frac{\left(\eta +\lambda\phi\right)^2}{12}\left(\int \frac{d^dk_{1}}{(2\pi)^d}\frac{d^dk_{2}}{(2\pi)^d}\frac{1}{k_{1}^2+m^{2}_{1}}\frac{1}{k_{2}^2+m^{2}_{1}}\frac{1}{(k_{1}+k_{2})^2+m^{2}_{1}}\right. \nonumber\\&& \left. +\left(\frac16 -\xi\right)R\int \frac{d^dk_{1}}{(2\pi)^d}\frac{d^dk_{2}}{(2\pi)^d}\frac{1}{k_{1}^2+m^{2}_{1}}\frac{1}{(k_{2}^2+m^{2}_{1})^2}\frac{1}{(k_{1}+k_{2})^2+m^{2}_{1}}\right. \nonumber\\&& \left. +\left(\frac16 -\xi\right)R\int \frac{d^dk_{1}}{(2\pi)^d}\frac{d^dk_{2}}{(2\pi)^d}\frac{1}{(k_{1}^2+m^{2}_{1})^2}\frac{1}{k_{2}^2+m^{2}_{1}}\frac{1}{(k_{1}+k_{2})^2+m^{2}_{1}}\right. \nonumber\\&& \left. +\left(\frac16 -\xi\right)R\int \frac{d^dk_{1}}{(2\pi)^d}\frac{d^dk_{2}}{(2\pi)^d}\frac{1}{k_{1}^2+m^{2}_{1}}\frac{1}{k_{2}^2+m^{2}_{1}}\frac{1}{((k_{1}+k_{2})^2+m^{2}_{1})^2}\right) = -\frac{\left(\eta +\lambda\phi\right)^2m^{2}_{1}}{2(4\pi)^4}\left(\frac{\mu^{-2\epsilon}}{\epsilon^2}+\frac{\mu^{-\epsilon}}{\epsilon}\left(\frac32+\psi(1)-\ln{\frac{m^{2}_{1}}{4\pi\mu^2}}\right) \right. \nonumber\\&& \left. +2\left(\ln\frac{4\pi\mu^2}{m^{2}_{1}}\right)^2 +2\left(\psi^{\prime}(1)+\psi^2(1)\right) +4 \psi(1)\ln\frac{4\pi\mu^2}{m^{2}_{1}}+\ln\frac{4\pi\mu^2}{m^{2}_{1}}+\psi(1)-\frac12\right)   +\frac{\left(\eta +\lambda\phi\right)^2}{2}\frac{\left(\frac16 -\xi\right)R}{(4\pi)^4}\left(\frac{\mu^{-2\epsilon}}{\epsilon^2}\right. \nonumber\\&& \left. +\frac{\mu^{-\epsilon}}{\epsilon}\left(\frac12+\psi(1)-\ln{\frac{m^{2}_{1}}{4\pi\mu^2}}\right)+2\left(\ln\frac{4\pi\mu^2}{m^{2}_{1}}\right)^2 +2\left(\psi^{\prime}(1)+\psi^2(1)\right) +4 \psi(1)\ln\frac{4\pi\mu^2}{m^{2}_{1}}+\ln\frac{4\pi\mu^2}{m^{2}_{1}}+\psi(1)-\frac12\right).
\label{w2}
\end{eqnarray}

\noindent
{\underline{\bf One loop counterterm contribution}}:\\

\noindent
The one loop counterterm contribution reads
\begin{eqnarray}
\frac12\left(\delta m^{2}+\frac12\delta \lambda \phi^2 +\delta \eta\phi+\delta \xi R\right)\times \int \frac{d^dk}{(2\pi)^d}\left(\frac{1}{k^2+m^{2}_{1}}+\frac{\left(\frac16 -\xi\right)R}{(k^2+m^{2}_{1})^2}\right),
\label{w3}
\end{eqnarray}
where $\delta m^{2}$, $\delta \lambda$ and  $\delta \xi$ can be seen in \ref{v29}. We next add the divergent terms coming from \ref{w1}, \ref{w2}, \ref{w3}, in order to find out the divergence of the effective potential
\begin{eqnarray}
&&\bold{Div}V_{\rm{eff}}^{2-{\rm loop}}= \frac{\mu^{-2\epsilon}}{2(4\pi)^4\epsilon^2}\left(2m^4\lambda +m^2\eta^2 +4m^2\eta\lambda\phi +\eta^3\phi +4m^2\lambda^2\phi^2 +\frac72\lambda\eta^2\phi^2 +3\eta\lambda^2\phi^3 +\frac34\lambda^3\phi^4 -2m^2\lambda \left(\frac16 -\xi\right)R \right.\nonumber\\&&\left. -\eta^2\left(\frac16 -\xi\right)R -4\eta\lambda\phi\left(\frac16 -\xi\right)R -\lambda^2\phi^2\left(\frac16 -\xi\right)R \right)\nonumber\\&& +\frac{\mu^{-\epsilon}}{2(4\pi)^4\epsilon}\left(\lambda\eta^2\phi^2\left(\frac52\psi(2)-\frac52\psi(1)-\frac{15}{4}\right)+2m^2\lambda^2\phi^2\left(\frac34 \psi(2)-\frac12\psi(1)-\frac34\right)+m^2\eta^2\left(\psi(2)-\psi(1)-\frac32\right)\right.\nonumber\\&&\left. +m^2\eta\lambda\phi\left(2\psi(2)-2\psi(1)-3\right)+\eta^3\phi\left(\psi(2)-\psi(1)-\frac32\right) +\frac12\lambda^3\phi^4\left(\psi(2)-\psi(1) -\frac32\right)\right.\nonumber\\&&\left. +\eta\lambda^2\phi^3\left(2\psi(2)-2\psi(1)-3\right)+\frac{\eta^2}{2}\left(\frac16 -\xi\right)R +\eta\lambda\phi\left(\frac16 -\xi\right)R+\frac{\lambda^2\phi^2}{2} \left(\frac16 -\xi\right)R    \right).
\label{w8}
\end{eqnarray}
The above equation leads to the two loop counterterms
\begin{eqnarray}
&&\delta\Lambda =-\frac{\mu^{-2\epsilon}\left(m^4\lambda +2m^2\eta^2\right)}{2(4\pi)^4\epsilon^2}-\frac{2\mu^{-\epsilon}m^2\eta^2\left(\psi(2)-\psi(1)-\frac32\right)}{2(4\pi)^4\epsilon}, \nonumber\\
&&\delta \kappa =\frac{\mu^{-2\epsilon}\left(2m^2 \lambda +\eta^2\right)}{2(4\pi)^4\epsilon^2}\left(\frac16 -\xi\right)-\frac{\eta^2\mu^{-\epsilon}}{4(4\pi)^4\epsilon}\left(\frac16 -\xi\right),\nonumber\\
&&\delta m^{2} =\frac{\left(4m^2\lambda^2 +\frac72\lambda\eta^2\right)\mu^{-2\epsilon}}{(4\pi)^4\epsilon^2}+\frac{2m^2\lambda^2\mu^{-\epsilon}}{(4\pi)^4\epsilon}\left(\frac34 \psi(2)-\frac12\psi(1)-\frac34\right)+\frac{\eta^2\lambda\mu^{-\epsilon}}{(4\pi)^4\epsilon}\left(\frac52\psi(2)-\frac52\psi(1)-\frac{15}{4}\right),\nonumber\\
&&\delta \xi =-\frac{\mu^{-2\epsilon}\lambda^2}{(4\pi)^4\epsilon^2}\left(\frac16 -\xi\right)+\frac{\lambda^2\mu^{-\epsilon}}{2(4\pi)^4\epsilon}\left(\frac16 -\xi\right),\nonumber\\
&&\delta \tau= \frac{\left(4m^2\eta\lambda +\eta^3 \right)\mu^{-2\epsilon}}{2(4\pi)^4\epsilon^2}+\frac{m^2\eta\lambda\mu^{-\epsilon}}{2(4\pi)^4\epsilon}\left(2\psi(2)-2\psi(1)-3\right)+\frac{\eta^3 \mu^{-\epsilon}}{2(4\pi)^4\epsilon}\left(\psi(2)-\psi(1)-\frac32\right),\nonumber\\
&&\delta \gamma = -\frac{4\eta\lambda\mu^{-2\epsilon}}{2(4\pi)^4\epsilon^2}\left(\frac16 -\xi\right)-\frac{\eta\lambda\mu^{-\epsilon}}{2(4\pi)^4\epsilon}\left(\frac16 -\xi\right), \nonumber\\
&&\delta \eta = \frac{9\mu^{-2\epsilon}\eta\lambda^2}{(4\pi)^4\epsilon^2}+\frac{3\mu^{-\epsilon}\eta\lambda^2}{(4\pi)^4\epsilon}\left(2\psi(2)-2\psi(1)-3\right),\nonumber\\
&&\delta \lambda = \frac{9\lambda^3\mu^{-2\epsilon}}{(4\pi)^4\epsilon^2}+\frac{6\lambda^3\mu^{-\epsilon}}{(4\pi)^4\epsilon}\left(\psi(2)-\psi(1)-\frac32\right).
\label{w17}
\end{eqnarray}

\ref{w17} renormalises the effective potential completely, giving us its finite parts coming from the double bubble and the sunset diagrams 
\begin{eqnarray}
&&V_{\rm{eff}}^{2-{\rm loop}} =-\frac{\lambda m^{4}_{1}}{2(4\pi)^4}\left(-\psi(2)\ln{\frac{m^{2}_{1}}{4\pi\mu^2}} +\frac12\left(\ln{\frac{m^{2}_{1}}{4\pi\mu^2}}\right)^2 +\frac14\left(\frac{\pi^2}{3}+2\psi^2(2)-\psi^{\prime}(2)\right)\right) \nonumber\\&&  + \frac{\lambda\left(\frac16 -\xi\right)Rm^{2}_{1}}{4(4\pi)^4}\left(2\left(\ln{\frac{m^{2}_{1}}{4\pi\mu^2}}\right)^2 +\psi(1)\psi(2)+\frac12\left(\frac{\pi^2}{3}+\psi^2(2)-\psi^{\prime}(2)\right) +\frac12\left(\frac{\pi^2}{3}+\psi^2(1)-\psi^{\prime}(1)\right)\right. \nonumber\\&& \left. +2\ln{\frac{4\pi\mu^2}{m^{2}_{1}}}\left(\psi(1)+\psi(2)\right)\right)-\frac{\left(\eta+\lambda\phi \right)^2 m^{2}_{1}}{2(4\pi)^4}\left(2\left(\ln\frac{4\pi\mu^2}{m^{2}_{1}}\right)^2 +2\left(\psi^{\prime}(1)+\psi^2(1)\right)+4 \psi(1)\ln\frac{4\pi\mu^2}{m^{2}_{1}}\right. \nonumber\\&& \left. +\ln\frac{4\pi\mu^2}{m^{2}_{1}}+\psi(1)-\frac12 \right)+\frac{\left(\eta+\lambda\phi \right)^2\left(\frac16 -\xi\right)R}{2(4\pi)^4}\left(2\left(\ln\frac{4\pi\mu^2}{m^{2}_{1}}\right)^2 +2\left(\psi^{\prime}(1)+\psi^2(1)\right)+4 \psi(1)\ln\frac{4\pi\mu^2}{m^{2}_{1}}\right. \nonumber\\&& \left. +\ln\frac{4\pi\mu^2}{m^{2}_{1}}+\psi(1)-\frac12 \right).
\label{w18}
\end{eqnarray}

\begin{figure}[H]
\begin{center}
\end{center}
\centering
\begin{subfigure}{.4\textwidth}
  \centering
  \includegraphics[scale=0.24]{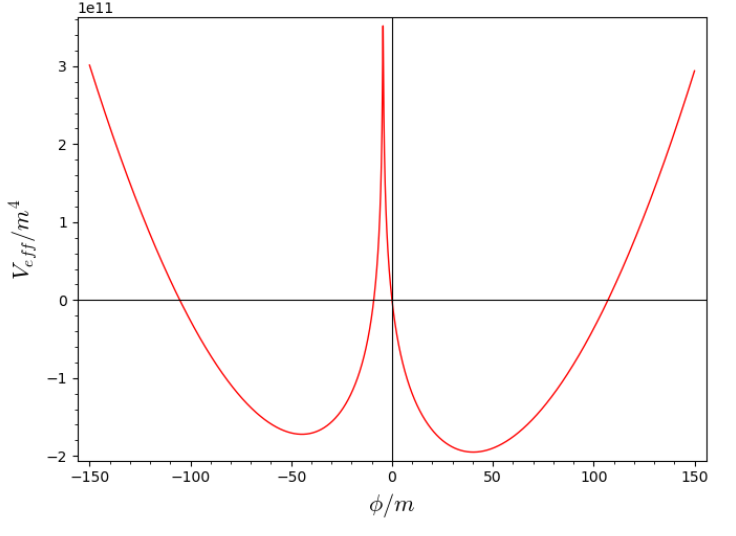}
  \caption{ $\eta\phi^3+\lambda\phi^4$ interaction }
  \label{fig:sub15}
\end{subfigure}%
\begin{subfigure}{.7\textwidth}
  \centering
  \includegraphics[scale=0.24]{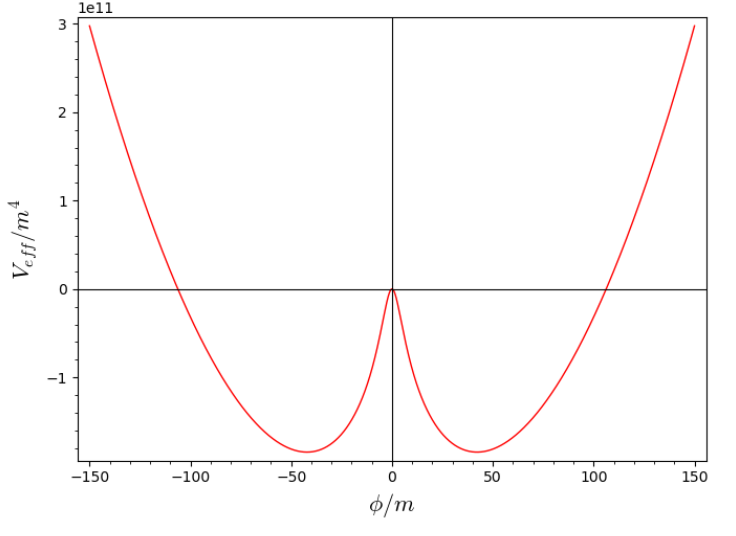}
  \caption{ $\lambda\phi^4$ interaction}
  \label{fig:sub16}
\end{subfigure}
\caption{\it \small The two loop double bubble effective potential, \ref{w18'}, at vanishing temperature in the de Sitter spacetime for values $\bar{\eta}=0.447, \ \lambda = 0.1, \ \xi =0.125,\ \Lambda=30{\rm GeV}^{2}$. Note that for the quartic self interaction, the symmetry breaking feature remains intact compared to that of one loop, \ref{fig:test4}.  As earlier, our renormalisation scheme ensures that there are no $\phi$-independent terms in the effective potential.}
\label{fig:test11}
\end{figure}

As we mentioned at the beginning of this section, the ${\cal O}(\lambda)$ double bubble diagram should make the leading contribution compared to the sunset which is quadratic in the couplings. Thus if we consider only the first, we have the two loop effective potential 
\begin{eqnarray}
&&V_{\rm{eff}}^{2-{\rm loop}}=V_{\rm{eff}}^{1-{\rm loop}}-\frac{\lambda m^{4}_{1}}{2(4\pi)^4}\left(-\psi(2)\ln{\frac{m^{2}_{1}}{4\pi\mu^2}} +\frac12\left(\ln{\frac{m^{2}_{1}}{4\pi\mu^2}}\right)^2 +\frac14\left(\frac{\pi^2}{3}+2\psi^2(2)-\psi^{\prime}(2)\right)\right) \nonumber\\&&  + \frac{\lambda\left(\frac16 -\xi\right)Rm^{2}_{1}}{4(4\pi)^4}\left(2\left(\ln{\frac{m^{2}_{1}}{4\pi\mu^2}}\right)^2 +\psi(1)\psi(2)+\frac12\left(\frac{\pi^2}{3}+\psi^2(2)-\psi^{\prime}(2)\right) +\frac12\left(\frac{\pi^2}{3}+\psi^2(1)-\psi^{\prime}(1)\right)\right. \nonumber\\&& \left. +2\ln{\frac{4\pi\mu^2}{m^{2}_{1}}}\left(\psi(1)+\psi(2)\right)\right) +{\cal O}(\lambda^2),
\label{w18'}
\end{eqnarray}
where the one loop effective potential is given by \ref{v30'}. However, we note that the one loop effective potential contains terms quadratic in the curvature as well, whereas in the two loop computation, we have retained terms only linear in the curvature. Hence we shall add with \ref{w18'} the second order in curvature contribution coming from the ${\cal O}(\lambda)$ double bubble, reading
\be
V_{{\rm eff},  {\cal O}(R^2)}^{2-{\rm loop}} =  -\frac{\lambda}{8(4\pi)^4}\left({2f_1}\left(\ln{\frac{m^2_{1}}{4\pi \mu^2}-\psi(2)}\right)+\left(\frac16-\xi\right)^2R^2\left(4\psi(1)\ln{\frac{4\pi\mu^2}{m_1^2}}+\frac32\left(\ln{\frac{4\pi\mu^2}{m_1^2}}\right)^2+\frac{\pi^2}{3}+2\psi^2(1)-\psi^{\prime}(1)\right)\right).  
\label{w18''}       
\ee
We shall not go into the detail of the derivation for the above. The above contribution only required some gravitational counterterms for the sake of renormalisation.  We have plotted \ref{w18'} {\it plus} \ref{w18''} in \ref{fig:test11} for the de Sitter spacetime. Note that compared to the zero temperature one loop result, \ref{v56}, there is no qualitatively new change. In particular for $\bar{\eta}=0$, the symmetry breaking feature remains intact. 
%

%%%%%%%%%%%%%%%%%%%
\subsection{Two loop at finite temperature}\label{S11}\
%%%%%%%%%%%%%%%%%%%%%%%
\noindent
Finally, let us come to finite temperature computations at two loop. In order to demonstrate  the feature of renormalisability clearly, instead of going to discretisation of the Euclidian $k^0$ as of \ref{S4}, we make the following decomposition (e.g.~\cite{Arjun, Ashok} and references therein)
\begin{eqnarray}
\frac{1}{k^2+m^2}\rightarrow\frac{1}{k^2+m^2}+2\pi n_{B}({\vert k^0 \vert})\delta(k^2-m^2) = \frac{1}{k^2+m^2}+n_1(\beta, k) \qquad ({\rm say}),
\label{w19}
\end{eqnarray}
and
\begin{eqnarray}
\frac{1}{\left(k^2+m^2\right)^2}\rightarrow\frac{1}{\left(k^2+m^2\right)^2}+n_2(\beta, k),
\label{w19'}
\end{eqnarray}
where $n_{B}({\vert k^0 \vert})=(\exp{\beta\vert k^0 \vert} - 1)^{-1}$ is the Bose distribution function.\\

\noindent
Thus the integrations will be like,
\begin{eqnarray}
\int \frac{d^dk}{(2\pi)^d}\frac{1}{k^2+m^2}\rightarrow \int \frac{d^dk}{(2\pi)^d}\frac{1}{k^2+m^2}+S_1(\beta)= -m^2\left(\frac{2\mu^{-\epsilon}}{\epsilon}+\psi(2)+\ln{\frac{4\pi\mu^2}{m^2}}\right)+S_1(\beta),
\label{w20}
\end{eqnarray}
where $S_1(\beta)$ is temperature dependent and finite. Likewise
\begin{eqnarray}
\int \frac{d^dk}{(2\pi)^d}\frac{1}{(k^2+m^2)^2}\rightarrow \int \frac{d^dk}{(2\pi)^d}\frac{1}{(k^2+m^2)^2}+S_2(\beta)=\left(\frac{2\mu^{-\epsilon}}{\epsilon}+\psi(1)+\ln{\frac{4\pi\mu^2}{m^2}}\right)+S_2(\beta).
\label{w21}
\end{eqnarray}

\noindent
{\underline{\bf Double bubble at finite temperature}}:\\

\noindent
The double bubble integral at finite temperature and up to ${\cal O}(R)$ reads
\begin{eqnarray}
&&-\frac{\lambda}{8}\left(\int \frac{d^dk}{(2\pi)^d}\frac{1}{k^2+m^{2}_{1}}+S_1(\beta)+\left(\frac16 -\xi\right)R\left(\int \frac{d^dk}{(2\pi)^d}\frac{1}{(k^2+m^{2}_{1})^2}+S_2(\beta)\right)\right)^2
\nonumber\\&& =-\frac{\lambda}{8}\left(\left(\int \frac{d^dk}{(2\pi)^d}\frac{1}{k^2+m^{2}_{1}}\right)^2+2\left(\frac16 -\xi\right)R \int \frac{d^dk}{(2\pi)^d}\frac{1}{k^2+m^{2}_{1}} \int \frac{d^dk}{(2\pi)^d}\frac{1}{(k^2+m^{2}_{1})^2} \right)\nonumber\\&&-\frac{\lambda}{8}\left(-\left(m^2+\eta\phi +\frac12\lambda\phi^2\right)\frac{4\mu^{-\epsilon}}{(4\pi)^2\epsilon}S_1(\beta)-\left(\frac16 -\xi\right)R\left(m^2+\eta\phi + \frac12 \lambda \phi^2\right)\frac{4\mu^{-\epsilon}}{(4\pi)^2\epsilon}S_2(\beta)+\left(\frac16 -\xi\right)R\frac{4\mu^{-\epsilon}}{(4\pi)^2\epsilon}S_1(\beta) \right)\nonumber\\&&+\rm{Finite\,\,terms }.
\label{w22}
\end{eqnarray}
Note that the first two terms are temperature independent divergent terms whereas the next three terms are temperature dependent divergent terms.\\
%%%%%%%%%%%%%%
%%
\pagebreak

\noindent
{\underline{\bf Sunset at finite temperature}}:\\

\noindent
The feynmann integral for Sunset diagram will be
\begin{eqnarray}
&&\frac{\left(\eta+\lambda\phi\right)^2}{12}\left(\int \frac{d^dk_{1}}{(2\pi)^d}\frac{d^dk_{2}}{(2\pi)^d}\left(\frac{1}{k_{1}^2+m^{2}_{1}}+n_1(\beta,k_1)\right)\left(\frac{1}{k_{2}^2+m^{2}_{1}}+n_1(\beta ,k_2)\right)  \left(\frac{1}{(k_{1}+k_{2})^2+m^{2}_{1}}+n_1(\beta ,k_1+k_2)\right)\right. \nonumber\\&& \left. +\left(\frac16 -\xi\right)R\int \frac{d^dk_{1}}{(2\pi)^d}\frac{d^dk_{2}}{(2\pi)^d}\left(\frac{1}{k_{1}^2+m^{2}_{1}}+n_1(\beta ,k_1)\right)\left(\frac{1}{(k_{2}^2+m^{2}_{1})^2}+n_2(\beta, k_2)\right)\left(\frac{1}{(k_{1}+k_{2})^2+m^{2}_{1}}+n_1(\beta ,k_1+k_2)\right)\right. \nonumber\\&& \left. +\left(\frac16 -\xi\right)R\int \frac{d^dk_{1}}{(2\pi)^d}\frac{d^dk_{2}}{(2\pi)^d}\left(\frac{1}{\left(k_{1}^2+m^{2}_{1}\right)^2}+n_2(\beta ,k_1)\right) \left(\frac{1}{k_{2}^2+m^{2}_{1}}+n_1(\beta ,k_2)\right)\left(\frac{1}{(k_{1}+k_{2})^2+m^{2}_{1}}+n_1(\beta ,k_1+k_2)\right)\right. \nonumber\\&& \left. +\left(\frac16 -\xi\right)R\int \frac{d^dk_{1}}{(2\pi)^d}\frac{d^dk_{2}}{(2\pi)^d}\left(\frac{1}{k_{1}^2+m^{2}_{1}}+n_1(\beta ,k_1)\right) \left(\frac{1}{k_{2}^2+m^{2}_{1}}+n_1(\beta ,k_2)\right)\left(\frac{1}{\left((k_{1}+k_{2})^2+m^{2}_{1}\right)^2}+n_2(\beta ,k_1+k_2)\right)\right)\nonumber\\&&=\frac{\left(\eta+\lambda\phi\right)^2}{12}\left(\int \frac{d^dk_{1}}{(2\pi)^d}\frac{d^dk_{2}}{(2\pi)^d}\frac{1}{k_{1}^2+m^{2}_{1}}\frac{1}{k_{2}^2+m^{2}_{1}}\frac{1}{(k_{1}+k_{2})^2+m^{2}_{1}}\right. \nonumber\\&& \left. +\left(\frac16 -\xi\right)R\int \frac{d^dk_{1}}{(2\pi)^d}\frac{d^dk_{2}}{(2\pi)^d}\frac{1}{k_{1}^2+m^{2}_{1}}\frac{1}{(k_{2}^2+m^{2}_{1})^2}\frac{1}{(k_{1}+k_{2})^2+m^{2}_{1}}\right. \nonumber\\&& \left. +\left(\frac16 -\xi\right)R\int \frac{d^dk_{1}}{(2\pi)^d}\frac{d^dk_{2}}{(2\pi)^d}\frac{1}{(k_{1}^2+m^{2}_{1})^2}\frac{1}{k_{2}^2+m^{2}_{1}}\frac{1}{(k_{1}+k_{2})^2+m^{2}_{1}}\right. \nonumber\\&& \left. +\left(\frac16 -\xi\right)R\int \frac{d^dk_{1}}{(2\pi)^d}\frac{d^dk_{2}}{(2\pi)^d}\frac{1}{k_{1}^2+m^{2}_{1}}\frac{1}{k_{2}^2+m^{2}_{1}}\frac{1}{((k_{1}+k_{2})^2+m^{2}_{1})^2}\right) \nonumber\\&& +\frac{\left(\eta+\lambda\phi\right)^2}{12}\left(\frac{6\mu^{-\epsilon}}{\epsilon}S_1(\beta)+\left(\frac16 -\xi\right)R\frac{6\mu^{-\epsilon}}{\epsilon}S_2(\beta)\right)+\rm{Finite\,\, terms}.
\label{w23}
\end{eqnarray}
The first four terms in the final expression are temperature independent divergent terms whereas next two terms are temperature dependent divergent terms.\\

\noindent
{\underline{\bf One loop counterterm contribution at finite temperature}}:\\

\noindent
The relevant integral reads
\begin{eqnarray}
&&\frac12\left(\delta m^{2}+\frac12\delta \lambda \phi^2 +\delta\eta\phi +\delta \xi R\right)\left(\int \frac{d^dk}{(2\pi)^d}\left(\frac{1}{k^2+m^{2}_{1}}+\frac{\left(\frac16 -\xi\right)R}{(k^2+m^{2}_{1})^2}\right)+S_1(\beta)+\left(\frac16 -\xi\right) R S_2(\beta)\right) \nonumber\\&&  =\frac12\left(\delta m^{2}+\frac12\delta \lambda \phi^2 +\delta\eta\phi +\delta \xi R\right)\left(\int \frac{d^dk}{(2\pi)^d}\left(\frac{1}{k^2+m^{2}_{1}}+\frac{\left(\frac16 -\xi\right)R}{(k^2+m^{2}_{1})^2}\right)\right) \nonumber\\&&  +\frac{\mu^{-\epsilon}}{2(4\pi)^2\epsilon}\left(-m^2\lambda -\eta^2 -\frac32\lambda^2\phi^2 -3\eta\lambda+\left(\frac16 -\xi\right)\lambda R\right)\left(S_1(\beta)+ \left(\frac16 -\xi\right) R S_2(\beta)\right).
\label{w24}
\end{eqnarray}
We now collect and add all the divergent terms from \ref{w22}, \ref{w23} and \ref{w24}.  We obtain after some calculations,
\begin{eqnarray}
&&{\rm Div}V_{\rm{eff},\beta}^{2-{\rm loop}}=-\frac{\lambda}{8}\left(\left(\int \frac{d^dk}{(2\pi)^d}\frac{1}{k^2+m^{2}_{1}}\right)^2+2\left(\frac16 -\xi\right)R \int \frac{d^dk}{(2\pi)^d}\frac{1}{k^2+m^{2}_{1}} \int \frac{d^dk}{(2\pi)^d}\frac{1}{(k^2+m^{2}_{1})^2} \right)\nonumber\\&& +\frac{\left(\eta+\lambda\phi\right)^2}{12}\left(\int \frac{d^dk_{1}}{(2\pi)^d}\frac{d^dk_{2}}{(2\pi)^d}\frac{1}{k_{1}^2+m^{2}_{1}}\frac{1}{k_{2}^2+m^{2}_{1}}\frac{1}{(k_{1}+k_{2})^2+m^{2}_{1}}\right. \nonumber\\&& \left. +\left(\frac16 -\xi\right)R\int \frac{d^dk_{1}}{(2\pi)^d}\frac{d^dk_{2}}{(2\pi)^d}\frac{1}{k_{1}^2+m^{2}_{1}}\frac{1}{(k_{2}^2+m^{2}_{1})^2}\frac{1}{(k_{1}+k_{2})^2+m^{2}_{1}}\right. \nonumber\\&& \left. +\left(\frac16 -\xi\right)R\int \frac{d^dk_{1}}{(2\pi)^d}\frac{d^dk_{2}}{(2\pi)^d}\frac{1}{(k_{1}^2+m^{2}_{1})^2}\frac{1}{k_{2}^2+m^{2}_{1}}\frac{1}{(k_{1}+k_{2})^2+m^{2}_{1}}\right. \nonumber\\&& \left. +\left(\frac16 -\xi\right)R\int \frac{d^dk_{1}}{(2\pi)^d}\frac{d^dk_{2}}{(2\pi)^d}\frac{1}{k_{1}^2+m^{2}_{1}}\frac{1}{k_{2}^2+m^{2}_{1}}\frac{1}{((k_{1}+k_{2})^2+m^{2}_{1})^2}\right),
\label{w25}
\end{eqnarray}
which is temperature independent, showing renormalisation is independent of the temperature. This serves as a consistency check of our calculations.\\
%%%%%%%%%%%%%%%%%%%%%%%%%%
%%

\noindent
After renormalisation, we can compute the finite part of the two loop effective potential at finite temperature. We wish to quote below only the result of the double bubble diagram we have found,
\begin{eqnarray}
&&V^{(2)}_{\rm{db} ,\beta} =V^{(2)}_{\rm{db}}+\frac{1}{576\beta^4}+\frac{m_1^2}{64\pi^2\beta^2}+\frac{m^{8}_{1}\beta^4}{65536\pi^8}\zeta^2(3)-\frac{m_1}{192\pi\beta^3}+\frac{m^{4}_{1}}{256\pi^4}\left(\ln\frac{m_{1}\beta}{2\pi}-\psi(1)\right)^2-\frac{m^{2}_{1}}{16\pi^2}\left(\ln\frac{m_{1}\beta}{2\pi} -\psi(1)\right)\left(\frac{1}{24\beta^2}\right. \nonumber\\&& \left. -\frac{m_1}{8\pi \beta} +\frac{m_{1}^{4} \beta^2}{256\pi^4}\zeta(3) - \frac{m_{1}^{6} \beta^4}{(32)(64)\pi^6}\zeta(5)+ \frac{15m_{1}^{8} \beta^6}{(96)(256)\pi^8}\zeta(7)\right)+\frac{m_{1}^{4}\zeta(3)}{6144\pi^4}-\frac{m_{1}^{5}\beta\zeta(3)}{2048\pi^5}-\frac{m_{1}^{6}\beta^2\zeta(5)}{49152\pi^6}+\frac{15m_{1}^{8}\beta^4\zeta(7)}{589824\pi^8} \nonumber\\&&  +\frac{m_{1}^{7}\beta^3\zeta(5)}{16384\pi^7}-\frac{15m_{1}^{7}\beta^3\zeta(7)}{196608\pi^9}-\frac{m_{1}^{10}\beta^6\zeta(3)\zeta(5)}{524288\pi^{10}}+2\left(\frac16 -\xi\right)R\left(\frac{1}{384\pi\beta m_1}-\frac{1}{128\pi^2\beta^2}\right. \nonumber\\&& \left. -\frac{m^{2}_{1}}{256\pi^4}\left(\ln\frac{m_{1}\beta}{2\pi} -\psi(1)\right)^2 +\frac{m^{2}_{1}}{16\pi^2}\left(\ln\frac{m_{1}\beta}{2\pi} -\psi(1)\right)\left(\frac{1}{16\pi\beta m_1}- \frac{m_{1}^{2} \beta^2}{128\pi^4}\zeta(3)+\frac{3m_{1}^{4} \beta^4}{2048\pi^6}\zeta(5) -\frac{15 m_{1}^{6} \beta^6}{3072\pi^8}\zeta(7)\right) \right. \nonumber\\&& \left. +\frac{1}{16\pi^2}\left(\ln\frac{m_{1}\beta}{2\pi} -\psi(1)\right)\left(\frac{1}{24\beta^2}-\frac{m_1}{8\pi\beta}+\frac{m_{1}^{4} \beta^2}{256\pi^4}\zeta(3) - \frac{m_{1}^{6} \beta^4}{(32)(64)\pi^6}\zeta(5) + \frac{15m_{1}^{8} \beta^6}{(96)(256)\pi^8}\zeta(7)\right)-\frac{m_{1}^{2}\zeta(3)}{3072\pi^4}\right. \nonumber\\&& \left. 
+\frac{m_{1}^{3}\beta\zeta(3)}{4096\pi^5}+\frac{m_{1}^{3}\beta\zeta(3)}{1024\pi^5}+\frac{3m_{1}^{4}\beta^2\zeta(5)}{49152\pi^6} -\frac{m_{1}^{5}\beta^3\zeta(5)}{32768\pi^7}-\frac{3m_{1}^{5}\beta^3\zeta(5)}{16384\pi^7}-\frac{m_{1}^{6}\beta^4\zeta(3)^2}{32768\pi^8}\right. \nonumber\\&& \left. -\frac{15m_{1}^{6}\beta^4\zeta(7)}{3072\pi^8}+\frac{15m_{1}^{7}\beta^5\zeta(7)}{24576\pi^9}-\frac{15m_{1}^{7}\beta^5\zeta(7)}{393216\pi^9} +\frac{3m_{1}^{8}\beta^6\zeta(3)\zeta(5)}{524288\pi^{10}}+\frac{m_{1}^{8}\beta^6\zeta(3)\zeta(5)}{262144\pi^{10}}\right) +{\cal O}(R^2),
\label{w26}
\end{eqnarray}
where $V^{(2)}_{\rm{db}}$ is the corresponding  zero temperature part and the cross terms of zero and finite temperature.
Inclusion of ${\cal O}(R^2)$ terms to the above result is easy, but we shall not go into such computation here.

%%%%%%%%%%%%%%%%%%
\section{Conclusion}\label{concl}
%%%%%%%%%%%%%%%

\noindent
Let us now summarise our work. In this paper we have considered the problem of the effective potential at zero and finite temperatures in a general curved spacetime. We have retained terms up to quadratic order in the curvature in the expansion of the Feynman propagator. We have focused upon the spontaneous symmetry breaking, its restoration and   phase transition phenomenon.  As we have emphasised earlier as well, one of our objectives was  to see whether the quadratic in curvature terms can bring in some qualitatively new physical  effect.

We have computed in \ref{S3},
the one loop effective potential at zero temperature, up to the quadratic order of the spacetime curvature for quartic and cubic self interactions. In \ref{S4}, we extended these results at finite temperature, with high temperature approximation.   We discuss spontaneous symmetry breaking, its restoration and phase transition at zero and finite temperatures in \ref{S6}, \ref{S8}. We next generalise some of these results to two loop in \ref{S9}. In particular, the contribution from the ${\cal O}(\lambda)$ two loop double bubble diagram under local approximation to the effective potential, at finite temperature  has been explicitly presented, \ref{S11}. For the de Sitter spacetime in particular, we have shown that we can have symmetry breaking for a scalar even with a positive rest mass squared {\it and} a positive non-minimal coupling, at zero temperature, at both one and two loop level.   This result remains valid for a very large range of renormalisation scale, and it cannot be achieved by linear curvature approximation.  We believe this result to be interesting in its own right. It is clear that due to symmetry breaking in such potential,  the new mass of the field will get contributions from the spacetime curvature, as well as non-minimal coupling. We have argued that such effect can be expected to be significant in an early universe scenario, where the density of the dark energy was supposed to be rather high compared to that of today.  \\

As we also noted earlier, the Schwinger-DeWitt expansion of the Feynman propagator, \ref{S2}, is essentially meant to capture the curvature effects into the ultraviolet of high energy quantum processes, such as the trace anomaly~\cite{Toms}. In other words, this method may not be suitable to probe, e.g., the deep infrared non-equilibrium effects relevant to the late time, super-Hubble scale during the primordial inflation~\cite{Prokopec:2002uw, Krotov:2010ma, Akhmedov:2013vka, Wang:2022mvv, Sadekov:2023ivd, Akhmedov:2024npw, Bhattacharya:2023yhx}   (also references therein), found by using the Schwinger-Keldysh closed time path formalism. Thus the results presented here can be thought of as sub-Hubble and short time scale, transient or even some initial phenomenon. An essential manifestation of this will be to note that the effective potential we found for the   de Sitter will also be valid for the static or any other patch, provided we erect our normal coordinate system in a small neighbourhood located much inside the cosmological event horizon. This is because the expansion coefficients of the Green function in \ref{v11}  consist only of curvature invariants, which are independent of any coordinate system.
Probing local Lorentz invariant physical processes in a curved spacetime background, to the best of our knowledge and understanding,  are important in their own right.
Also, after the occurrence of the SSB at short spacetime scales, the field acquires a new mass, which can affect its dynamics at large scales at late times. It may be interesting to look into the late time long wavelength  or stochastic behaviour of the background scalar field, the equation of motion of which can be found from the effective action corresponding to the effective potential.  

 In this connection we also note that, in a curved spacetime, the notion of temperature and thermal equilibrium might be ambiguous \cite{Tolman:1930ona}. This may correspond to, for example, the absence of spatial isotropy, the red or blueshift effects etc. However, suppose we confine ourselves to a small region of the spacetime and look for the leading departure from that of the flat spacetime. In that case, it seems that such issues are taken care of, at least approximately. Certainly by construction, the Schwinger-DeWitt expansion technique seems to be appropriate to address such local thermal issues in a curved background, e.g. \cite{Hu}.
 
   Also, as we noted earlier towards the end of \ref{S6}, that  the above results perhaps indicate that we should employ some  non-perturbative techniques in order to investigate further the symmetry breaking and phase transition phenomena in curved spacetimes. This not only means resummation of the coupling constant, but also at least some of the curvature terms appearing in the expansion of the propagator. This proposition is also motivated by the one loop self energy computation, \ref{g4}.  Note once again that such resummed formalism is not expected to capture the infrared physics, as the framework would still be based upon the local Lorentz invariance and four momenta. Instead, this attempts to resum the curvature effects in the local, ultraviolet phenomenon.  Finally, we also note that  the  $\phi^4$ theory is a toy model, and in order to see whether there is any phenomenological consequences of the results we obtain here, we must go to scalars coupled to non-Abelian gauge fields. While the present work can be considered as a first step in order to gain some initial insights, we hope to come back to the aforementioned issues sometimes soon in future publications.

%%%%%%%%%%%
\section*{Acknowledgement} The work of VN is supported by the junior research fellowship of University Grants Comission, Govt. of India (Ref. no.~211610068936).  He also wishes to thank B.~L.~Hu and I.~L.~Shapiro for useful communications. The authors would like to acknowldge anonymous referees for various useful comments.  
%%%%%%%%%%

%\bigskip
%\appendix
%\labelformat{section}{Appendix #1} 
%\section{The diagrams with purely non-local contributions but no non-vanishing 
%%%%%%%%%%%%%%%

\end{document}